\documentclass[%
reprint,
superscriptaddress,
nobibnotes,
amsmath,
amssymb,
aps,
prl,
]{revtex4-2}

\usepackage{graphicx}
\usepackage{dcolumn}
\usepackage{bm}
\usepackage[dvipsnames]{xcolor}
\usepackage{placeins}

\begin{document}

\title{Role of Local Structural Variation in X-ray Photoelectron Spectrum of Silicon Oxide Interfaces}

\author{Mikael Santonen}
\affiliation{University of Turku, Department of Physics and Astronomy, FI-20014 Turun yliopisto, Finland}
\author{Sari Granroth}
\affiliation{University of Turku, Department of Physics and Astronomy, FI-20014 Turun yliopisto, Finland}
\author{Johanna Laaksonen}
\affiliation{University of Turku, Department of Physics and Astronomy, FI-20014 Turun yliopisto, Finland}
\author{Pekka Laukkanen}
\affiliation{University of Turku, Department of Physics and Astronomy, FI-20014 Turun yliopisto, Finland}
\author{Johannes Niskanen}
\email{johannes.niskanen@utu.fi}
\affiliation{University of Turku, Department of Physics and Astronomy, FI-20014 Turun yliopisto, Finland}

\begin{abstract}
We show that the broad X‑ray photoelectron lines of silicon oxide on silicon arise from a continuous statistical distribution of core-level binding energies. Statistical simulations spanning compositions from Si to SiO$_2$ reproduce the full extent of this broadening, reaching $\sim$5 eV for SiO$_{1.0}$, in quantitative agreement with 0.23 nm layer-resolved spectra reconstructed from Ar$^+$ sputtering data. This continuous distribution blurs distinct spectral fingerprints of local structural motifs, thereby challenging conventional chemical state assignment in oxide X-ray photoelectron spectra.
\end{abstract}

\maketitle

\section*{}
The operation of many contemporary electronic devices relies on ultrathin SiO$_x$ films with physical properties less well characterized than those of bulk SiO$_2$. Even when using atomic-layer-deposited HfO$_2$ and Al$_2$O$_3$, incorporation of a SiO$_2$ interlayer between Si and the deposited film has been shown to improve the performance of many devices, {\it e.g.} by decreasing leakage currents \cite{Cheema2022}. Furthermore, new applications for thin SiO$_2$ layers are emerging. For example, the material is increasingly being used as part of the tunneling contacts in high-efficiency silicon solar cells \cite{Khler2018}. Monitoring the formation of SiO$_x$ on Si and characterizing the film require highly specific tools such as X-ray photoelectron spectroscopy (XPS). The short mean free path of the photoelectron gives XPS sensitivity only to the outermost nanometers of the sample, making the method well suited for studies of surfaces and interfaces in thin-film systems, particularly in the semiconductor industry and for SiO$_x$. As the oxide film thins, the proportion of substoichiometric SiO$_x$ increases, making its contribution to the measured XPS signal more substantial and its reliable interpretation increasingly important. The common interpretation for Si 2p spectra is that each averaged Si oxidation state yields a single core-level component as a fingerprint, shifted relative to the Si substrate peak: +1~eV (Si$^{1+}$), +2~eV (Si$^{2+}$), +3~eV (Si$^{3+}$) and +4~eV (Si$^{4+}$) \cite{Keister1999}. Furthermore, these features are often interpreted to imply a stack of intermediate oxide layers, with oxidation state increasing toward +IV in SiO$_2$. Our results obtained by XPS and high-performance computing reveal that these assumptions are flawed and can lead to incorrect conclusions about the properties of SiO$_x$. 
\par
\begin{figure}
    \centering
    \includegraphics[width=\columnwidth]{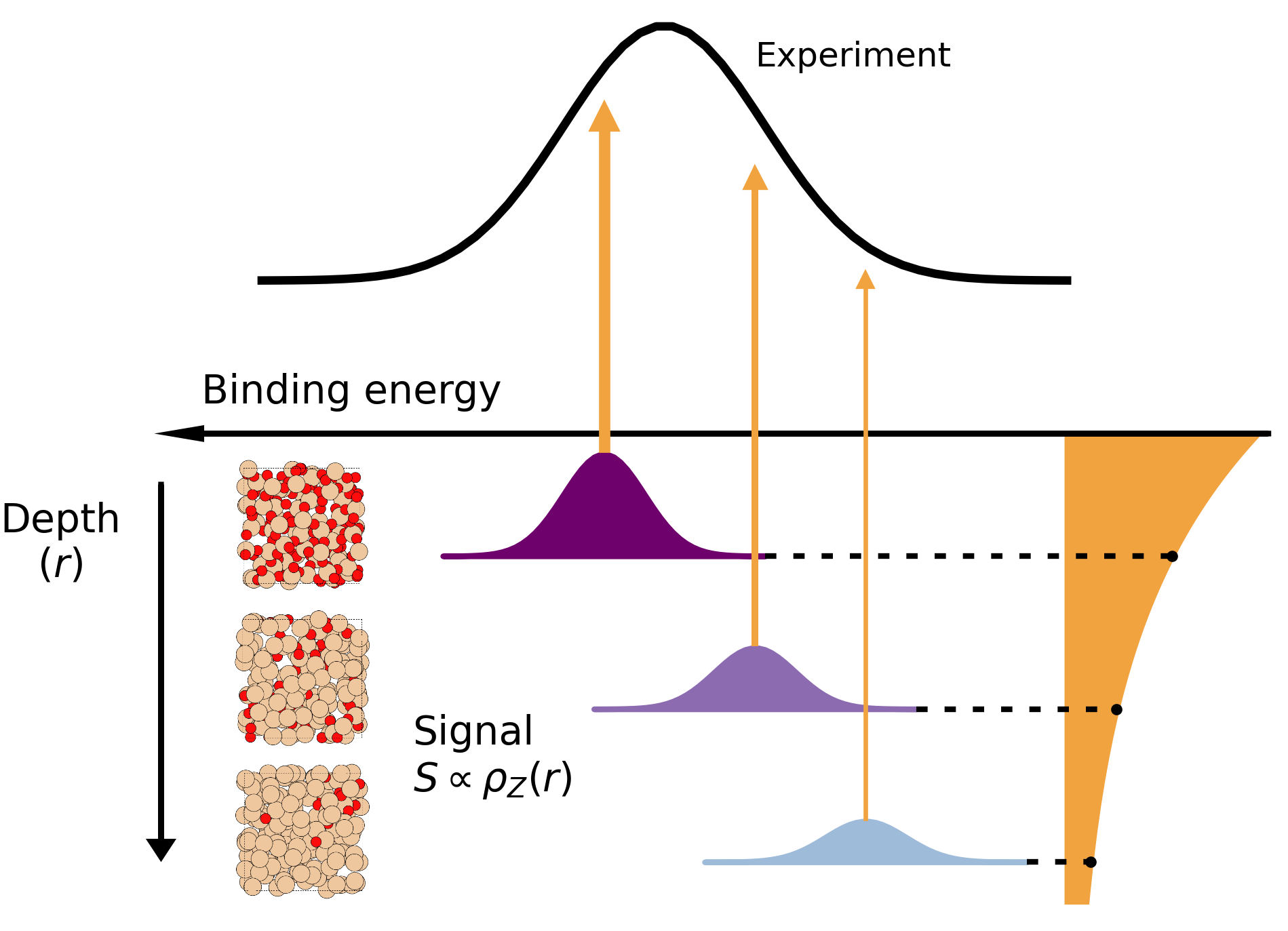}
    \caption{The build-up of an X-ray photoelectron spectrum in surface-oxidized silicon. The chemical shift for an atom depends on the local oxygen number density $\rho_\mathrm{O}(r)$, which decays towards bulk silicon. In case of oxygen, the native signal strength at depth $r$ is inherently proportional to $\rho_\mathrm{O}(r)$, whereas for silicon the respective number density $\rho_\mathrm{Si}(r)$ and native signal strength vary less. Attenuation due to inelastic electron scattering yields an exponentially decaying factor for the native signal in the depth-integrated spectrum. The mean free path $\lambda_\mathrm{ph}$ of the incident photon is assumed to be orders of magnitude longer than the mean free path $\lambda_\mathrm{el}$ of the photoelectron.}
    \label{fig:bigpicture}
\end{figure}
\par
The sensitivity of core-level spectra to structural characteristics varies strongly. For example, using data-driven analysis, spectra of liquids with thousands of degrees of freedom have been found to be governed by a few latent variables, which are not necessarily those assumed {\it a priori} \cite{Eronen2024b,Eronen2025}. In solids, additional characteristic phenomena influence the spectra. However, in terms of statistical variation, the situation is similar: if a large variety of local structures is present, their individual contributions to the recorded ensemble-averaged spectrum may vary significantly. As an example, in the liquid phase, a broadening of an electronvolt has been found to originate from thermal-motion-induced structural variation \cite{Ottosson2011, Niskanen2013, Löytynoja2014}, which renders an analysis using a single representative structure a potential oversimplification. In this work we challenge the motif-based view of oxidized silicon and explain the core-level photoelectron lines by structural diversity with a continuous influence on the core-level binding energies (BE). First, we reconstruct the XPS signal for individual layers in a silicon oxide interface from experimental XPS data at the O 1s and Si 2s edges, thereby avoiding spin-orbit splitting. We then investigate the variation of the respective BEs with local structure using statistical simulations. The experiment can be explained using typical assumptions for core-level binding energy calculations, provided that the aforementioned variation is accounted for. Finally, we determine the radius of spectral sensitivity to be 5~{\AA}.
\par
\begin{figure*}
    \centering
    \includegraphics[width=\textwidth]{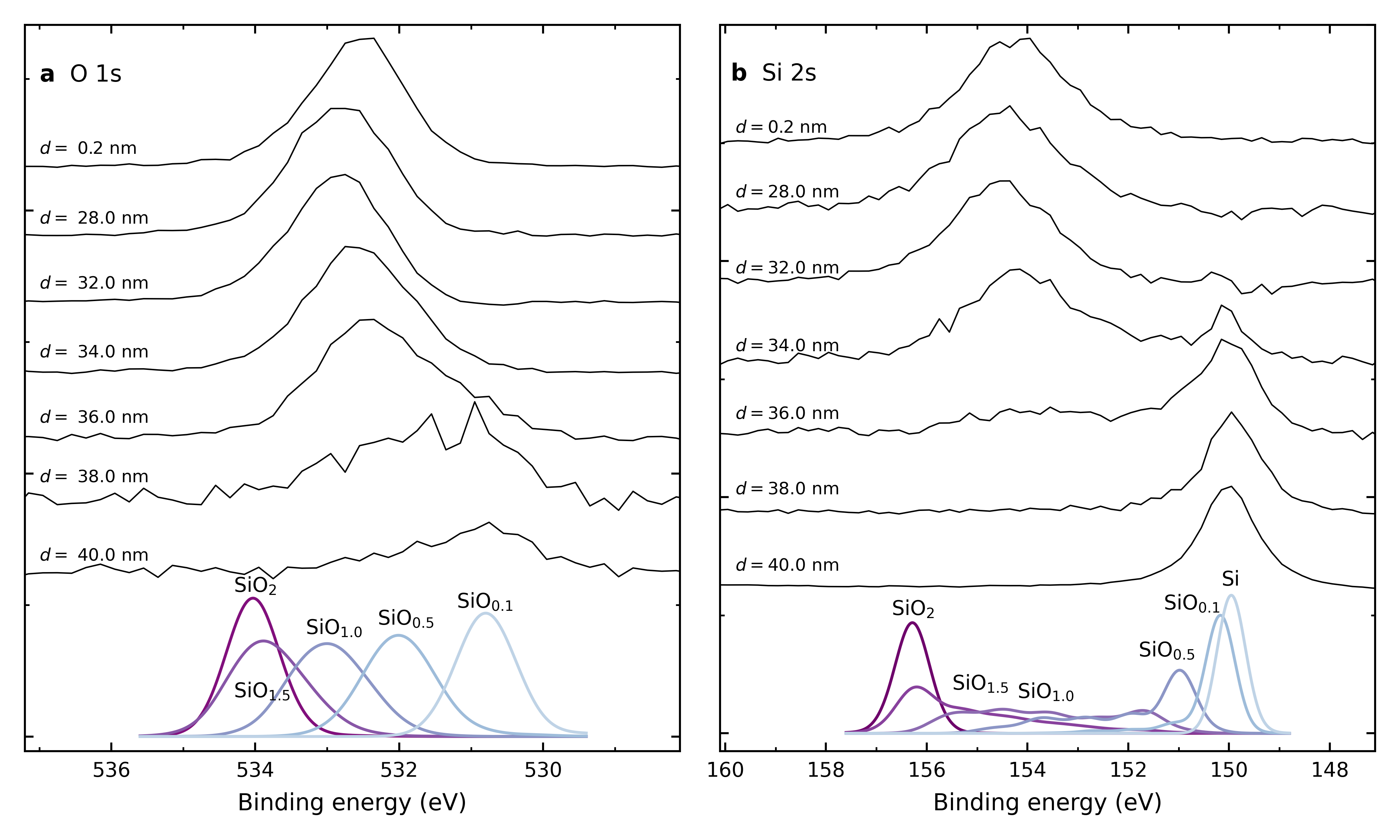}
    \caption{Processed layer-wise spectra for O 1s (a) and Si 2s (b) derived from observed raw X-ray photoelectron spectra for sputtered silicon oxide is shown on top with $d$ showing the sputtering depth of the topmost raw specrum in the reconstruction. The layer-wise spectra were binned with consecutive pairs to reduce noise. Simulated ensemble-averaged O 1s and Si 2s spectra for different compositions of SiO$_x$ and pure Si are shown in the bottom. The simulated spectra have been shifted by 3.6~eV (O 1s) and 12.0~eV (Si 2s) towards the experimental peaks. 5000 sites were calculated for each composition with sites from tens anneal runs for each. A weak plasmon loss peak from the Si 2p overlaps with the Si 2s peak (Fig. S9 in SM). The contribution, estimated from the Si 2s plasmon, is at most 3\% of the height of the Si 2s peak.}
    \label{fig:ensemble}
\end{figure*}
We recorded experimental data along the depth of a commercial Si(100) piece with an in-house grown 40 ± 2 nm thick SiO$_2$ layer repetitively with sputtering with a 300 eV monoatomic beam of Ar$^+$ and an etch time of 20 s per sputter cycle. The oxide layer was grown by thermal oxidation of Si(100) using an oven with flowing O$_2$, heated above 800\textdegree C during the oxidation. The thickness of the SiO$_2$ layer was measured with a Gaertner L116S ellipsometer. Prior to the measurements, we removed the organic residues from the sample with subsequent immersions in acetone and isopropanol for three minutes in conjunction with ultrasonication, followed by drying with N$_2$. We measured the raw XPS spectra using the Thermo Scientific Nexsa system with a pass energy of 30~eV and a kinetic energy step of 100~meV. For Ag 3d$_{5/2}$, the full width at half maximum (FWHM) of 0.64~eV was achieved. We focused a monochromated Al K$\alpha$ characteristic X-ray line to a spot with a diameter of roughly 300~$\mu$m, and counter-acted the resulting charging by dual-beam charge compensation.
\par
As illustrated in Figure \ref{fig:bigpicture}, the observed raw spectrum $\mathbf{s}_i^{\mathrm{(obs)}}$ after sputtering cycle $i$ is an integral of contributions from all sample below. We treat this integral as a sum over layers with thickness $\Delta r$, each attenuated by inelastic electron scattering with the mean free path $\lambda_\mathrm{el}$ due to the layers above it at the particular state of the process. Using the depth increase in a single sputtering cycle as $\Delta r$ yields an approximative expression in terms of the native spectra $\mathbf{s}_j$ of layers $j\geq i$ still in place:
\begin{equation}
\label{eq:effmodel}
\mathbf{s}_i^{\mathrm{(obs)}} = \sum_{j\geq i}\underbrace{\mathrm{e}^{-(j-i)\Delta r/\lambda_\mathrm{el}}}_{=:a_{ij}}\mathbf{s}_j.
\end{equation}
The equation can be cast in matrix form $\mathbf{S}^{\mathrm{(obs)}}=\mathbf{AS}$, where matrix $\mathbf{A}$ contains coefficients $a_{ij}$ as its elements, and matrices $\mathbf{S}^{\mathrm{(obs)}}$ and $\mathbf{S}$ contain the observed and the layer-wise spectra as their row vectors, respectively. Matrix $\mathbf{A}$ has the determinant of unity, and inspection of the inverse matrix $\mathbf{A}^{-1}$ reveals that the spectrum $\mathbf{s}_j$ of layer $j$ equals to the difference of the raw observations $j$ and $j+1$ after scaling the latter down by the factor $\mathrm{e}^{\Delta r/\lambda_\mathrm{el}}$ due to the attenuation caused by layer $j$. We assume the mean free path of the incident photons to be large and omit the related effects. The model considers an ideal layer-by-layer removal, supported by low ion energy \cite{Hofmann2005} and nearly stoichiometric preferential sputtering of SiO$_2$ \cite{Malherbe1986}. Validation of the method and the full experimental dataset are given in the Supplemental Material \cite{supplemental-arxiv} (Fig. S8, S15).
\par
Figure \ref{fig:ensemble} shows representative reconstructed layer-wise spectra $\mathbf{s}_j$ labeled with the etch time $t_e$ of the topmost raw spectrum. An etch rate of 0.01 nm/s ($\Delta r\approx0.2$ nm per sputter cycle), for which the estimation is presented in the Supplementary Material (Fig. S10), and $\lambda_\mathrm{el} = 3.5$ nm \cite{Tanuma1994} was used for the reconstruction. For both O 1s and Si 2s, the silicon oxide peak shifts toward lower binding energies with depth. The intermediate region is characterized by pronounced broadening and wide spread of binding energies. As we sputter the oxide layer away, the number density of oxygen decreases and the oxygen signal becomes weaker. As a result, the O 1s peak and the silicon oxide component of Si 2s disappear. With Si 2s, the pure silicon peak emerges at intermediate depths. It features a shoulder on the high BE side, which decays as oxygen is sputtered away.
\par
To interpret the experimental spectra, we carried out statistical simulations of core-level binding energies by calculating spectra from snapshots sampled from molecular dynamics (MD) runs. We simulated $\sim$15~Å periodic simulation cells of SiO$_x$ ($x$=0, 0.1, 0.5, 1.0, 1.5, and 2.0) using the Atomic Simulation Environment \cite{Larsen2017} and the MACE machine learning potential \cite{batatia2025}. The initial atomic positions were randomly generated for the different compositions such that the resulting box was roughly 15 Å and the density was a fixed 2.2 g/cm$^3$. After a local minimization, the system was melted with an NVT MD run using the Langevin thermostat at $T$ = 3000 K for 100 ps. The timestep for all MD runs was 1 fs. For the cooling stage, we switched to an NpT simulation (isotropic Martyna-Tobias-Klein dynamics) at 1 bar external pressure, which allowed the cell size to relax \cite{Cole2007}. Time constants for the temperature and pressure were 100 fs and 1000 fs, respectively. The annealing was done with the following schedule \cite{Vashishta1990}: the target temperature was reduced in steps of 500 K with 30 ps of MD equilibration at each step until reaching 300 K. The resulting cooling rate was on the order of 10$^{13}$ K/s. Because MD of the quenched solid does not sample phase space well, annealing runs were carried out as branches from the molten simulation endpoint at the interval of 3~ps, which exceeds the decay time of the structural autocorrelation function (Fig. S7 in SM). These extensive simulations were enabled by the ML interaction potential and graphics processing units. From tens branches for each system, we calculated a total of 5000 binding energies per system and XPS peak.
\par
We used density functional theory (DFT) implemented in the GPAW software \cite{Mortensen2024} v.22.1.0 and the Atomic Simulation Environment \cite{Larsen2017} to evaluate core-level binding energies, following Ref. \citenum{Eronen2025}. The calculations were carried out with the $\Delta$-SCF approach, Perdew-Burke-Ernzerhof (PBE) exchange-correlation functional \cite{Perdew1996}, periodic boundary conditions, $\Gamma$-point-only Brillouin-zone sampling and a plane-wave cutoff of 600~eV. The ground state and full core-hole atomic projector augmented wave (PAW) setups were generated using the GPAW implementation with its default values. To aid convergence, occupation smearing by the Fermi-Dirac distribution with a width 0.25~eV was used. For each case (O 1s, Si 2s), we assume that the core-level ionization cross section is the same 
for all of the snapshots. This is a reasonable approximation for photon energies well above the respective core-level ionization threshold. The data thus form stick spectra, which we broadened by convolution with a Gaussian function with a full-width at half maximum of 0.5~eV. The value was chosen to account for vibrational lineshape, lifetime broadening and instrumental effects, but we note that the exact value is not critical for our conclusions. The calculations used an effective core-potential for the hole state accounting for overall core relaxation and explicit reordering of valence. The raw $\Delta$-SCF binding energies were shifted to align with the deepest peak of our experiment, by 3.6 eV for O 1s and by 12.0 eV for Si 2s.
\par
We validated the simulations in numerous ways. For example, further geometry optimization within the simulation cell using the DFT method of spectrum calculations did not have a notable effect on the ensemble-averaged spectra. Likewise, the effect of the cell size was tested with a 2$\times$2$\times$2 supercell. The binding energy difference between sites in Si and SiO$_2$ was 5.8 eV in the 15 Å cell and 6.1 eV in the supercell. We assessed the sensitivity  to the exchange-correlation functional by comparing calculations with Local Density Approximation (LDA) \cite{Kohn1965} and PBE. We found that, while the absolute binding energies differed by about 0.7 eV compared to those from PBE, the binding energy shifts differed only by 0.01 eV. This suggests that the relative energies are not sensitive to the exchange-correlation functional in this case. Last, a good match for the binding energy shifts of ethyl-trifluoroacetate by Travnikova and coworkers \cite{Travnikova2012} was obtained using the procedure as shown in Supplementary Material (Fig. S1a).
\par
The calculations exhibit notable variation of BEs, as shown in Figure \ref{fig:ensemble}, which depicts the calculated ensemble-averaged spectra for each of the systems. These calculated spectra cover the experimentally observed range and explain the shifts observed in the experimental spectra. However, the method overestimates the shift in the calculated spectra. Due to this overestimation and the wide continuous spread of core-level binding energy values, we focus on
trends in this highly statistical issue avoiding non-unique, potentially misleading fits to the experiment. For both cases, the calculated peaks also shift clearly toward lower binding energies with decreasing $x$ in SiO$_x$.
The mechanism behind the shift to higher binding energies along increasing $x$ is two-fold, as shown in the SM section \textit{Population analysis}. For silicon, local electron population can vary several elementary charge units which gives rise to the effect. On the contrary, for the oxygen atoms the variation of local charge is an order of magnitude smaller, and the charge in the local neighborhood explains the binding energy shift. As this information is contained by the nuclear configuration and the total charge of the system, the binding energy at a given site is a function of them only via the solution of the respective electronic Schrödinger equation.
The variation in the intermediate region is explained by the intermediate oxygen number densities featuring a large amount of structural variation, which leads to a wide spread in the binding energies. In both cases, the binding energies of the intermediate stoichiometries span over several electronvolts, with overlap in the peaks. Most notably, SiO$_{1.0}$ features the largest variation of over 5 eV for the Si 2s spectrum.
In the O 1s spectra (Fig. \ref{fig:ensemble}a), the SiO$_2$ peak is notably narrower than the rest. This is consistent with the lower structural variation in the local atomistic structure of SiO$_2$. We found that most first-neighbor environments for SiO$_{1.5}$ are still similar to those of SiO$_2$, which explains the comparable peak centers. However, structural distortions and oxygen-deficient local environments lead to the increased variation to both high and low BE directions. 
The SiO$_{0.1}$ and SiO$_{0.5}$ peaks explain the previously mentioned shoulders, observed in the intermediate-depth experimental Si 2s spectra. 
\par
The low-oxygen limit of a suboxide is given by a single oxygen atom in a spectroscopically relevant neighborhood. At this limit some effects on Si 2s are expected, which we studied by calculations for a cell with only one oxygen atom and 149 silicon atoms, giving us an atomic oxygen concentration of 0.67\%. Most of the resulting Si 2s binding energies are within the range of variation for typical for pure Si. However, two silicon atoms show binding energies, that are clearly shifted by nearly 1 eV from the main peak center. The resulting spectrum is shown in the SM (Fig. S15).
\par
\begin{figure}
    \centering
    \includegraphics[width=\columnwidth]{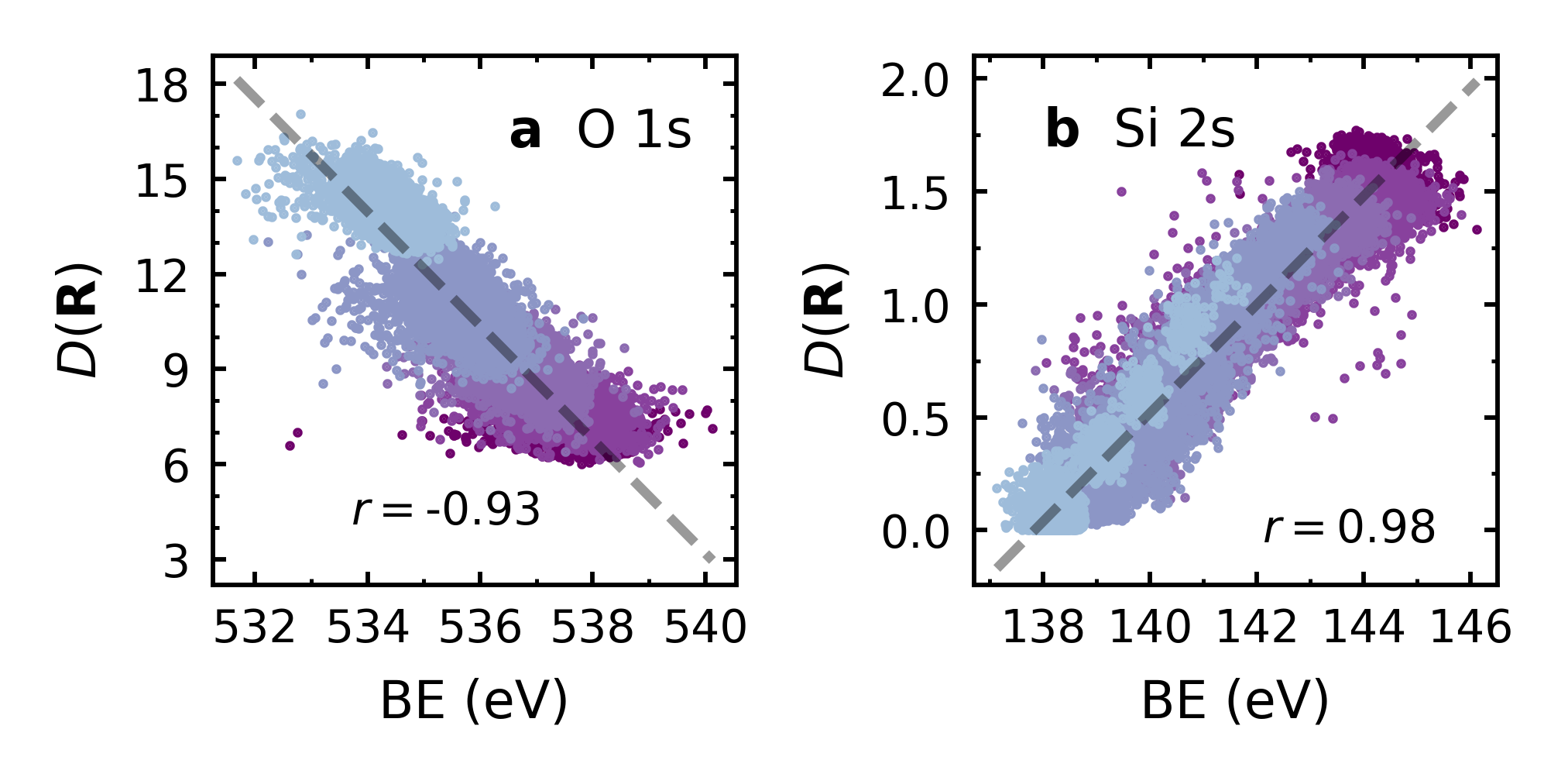}
    \caption{The dependence of the O 1s (a) and Si 2s (b) binding energies on the local atomistic environment descriptor $D(\mathbf{R})$ shows a clear correlation. The opposite trends originate from a different element being used in the respective descriptor. The colors correspond with the peaks in Figure \ref{fig:ensemble}. For details, see text.}
    \label{fig:desc}
\end{figure}

To further investigate the BE variation caused by that of the local atomistic structures, we calculated the dependence of the BE value on a simple atomistic-structure descriptor. We define this quantity $D(\mathbf{R})=\sum_i \exp(-r_i/\lambda)$, where the sum runs over all Si atoms for O 1s BE, and over all O atoms for Si 2s BE. The radial distance from the ionization site is denoted by $r_i$, and $\lambda$ is a decay constant chosen to maximize correlation, 3 {\AA} for O 1s BE and 1.1 {\AA} for Si 2s BE. The results are shown in Figure \ref{fig:desc}. 
We observe a clear correlation between the binding energy and the descriptor of the local atomistic structure. To give an example of the effect of the derived relation, we can estimate that an additional atom needs to be at least $\sim$5 {\AA} from the ionization site to contribute less than 0.1 eV to the binding energy shift. While the decay constants for the different cases differ, the estimated, spectrally relevant, environment is similarly sized for both: 5.2 {\AA} for O 1s BEs and 4.8 {\AA} for Si 2s BEs. 
These conclusions are consistent with the 2$\times$2$\times$2 supercell calculation, which already suggests that the sensitivity of the core-level shift originates from the local atomistic configuration at the site of ionization.
We tested various descriptors with different radial weightings. A $1 / r_i$-type form performed worse, suggesting that the screening effects, captured by the exponential form, are relevant here. We also compared summing over all atoms, only oxygen atoms, and only silicon atoms. The descriptor performed best when summing was done over only the silicon atoms in the O 1s BE case and only oxygen atoms in the Si 2s BE case. The coordination of the first-neighbor atoms is expectedly a dominant factor and the best performing descriptor reflects this. However, as the descriptor also shows, atoms beyond the first coordination shell still measurably affect the resulting binding energies in a continuous fashion. This result is inconsistent with modeling the spectra with a few discrete local motifs.
\par
The vibrational lineshape in XPS may be misinterpreted as a fingerprint of a separate chemical state \cite{Bancroft2009}.
While our analysis focuses on Si 2s, the shifts are analogous to those of Si 2p, where the signal between the bulk Si and SiO$_2$ peaks have commonly been fitted and assigned to three intermediate oxidation states Si$^{1+}$, Si$^{2+}$, Si$^{3+}$ corresponding with silicon atoms with bonds to 1, 2, and 3 oxygen atoms \cite{Himpsel1988, Keister1999, Stegemann2017}. The interpretation has been questioned \cite{BanaszakHoll1994}, and Zhang et al. \cite{Zhang1998}, for example, provide experimental evidence that the nearest-neighbor coordination is insufficient in explaining the intermediate signal and deeper consideration of the local atomistic environment is required. Our results support this conclusion.
\FloatBarrier
The broad spread of the simulated core-level binding energies, both between and within SiO$_x$ systems with varying $x$, explains the observed X-ray photoelectron spectra. These results highlight the difference between the macroscopic oxygen number density and the spectroscopically probed oxygen density, with an estimated spectroscopically relevant radius of around 5 {\AA}. 
The broad peak shapes observed in the spectra are not indicative of any single structural motif, but rather result from averaging over local mean oxygen density $x$ and internal statistical variation within local structures of a given $x$. These statistical mechanisms of variation render fitting procedures qualitative and likely unsuitable for identifying local structural motifs in SiO$_x$ with intermediate $x$. Moreover, even though adding oxygen atoms to a local neighborhood increases the core-level binding energies therein, the statistical distribution of such local environments blurs the distribution of the values so that individual distinct oxidation states can not be identified from the X-ray photoelectron spectrum as separate features.
\section*{}
\textit{Acknowledgements} --- 
J.N. acknowledges funding by Research council of Finland via the academy project grant 367978. The authors acknowledge CSC – IT Center for Science, Finland, for extremely generous computational resources and thank Dr. O. Krejci for introduction to the MACE potential. We thank Dr. E. A. Eronen for discussions about spectral informatics. 
\section*{}
\textit{Contributions} --- M.S. simulations, analysis, writing; S.G. experiment, writing; J.L. experiment, writing; P.L. research design, writing; J.N. lead research design, simulations, writing, funding.

\section*{}
\textit{Data availability} --- Data and scripts supporting the findings of this work are openly available \cite{dataset}.

\bibliography{bibliography}

\onecolumngrid
\clearpage

\section*{Supplementary Material}

\section{Validation of the binding energy calculations}
\begin{figure}[h]
\includegraphics[width=0.49\textwidth]{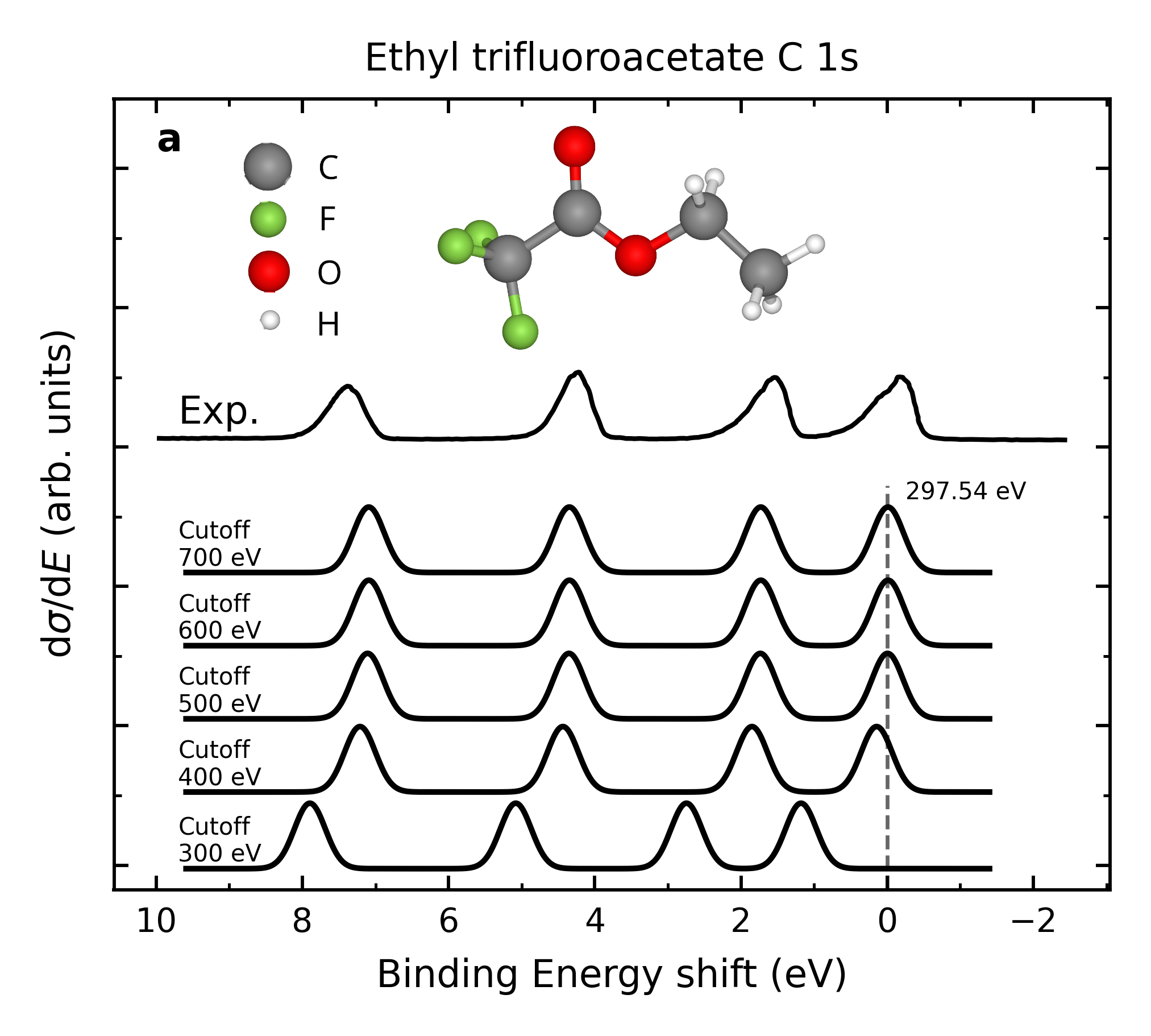}
\includegraphics[width=0.49\textwidth]{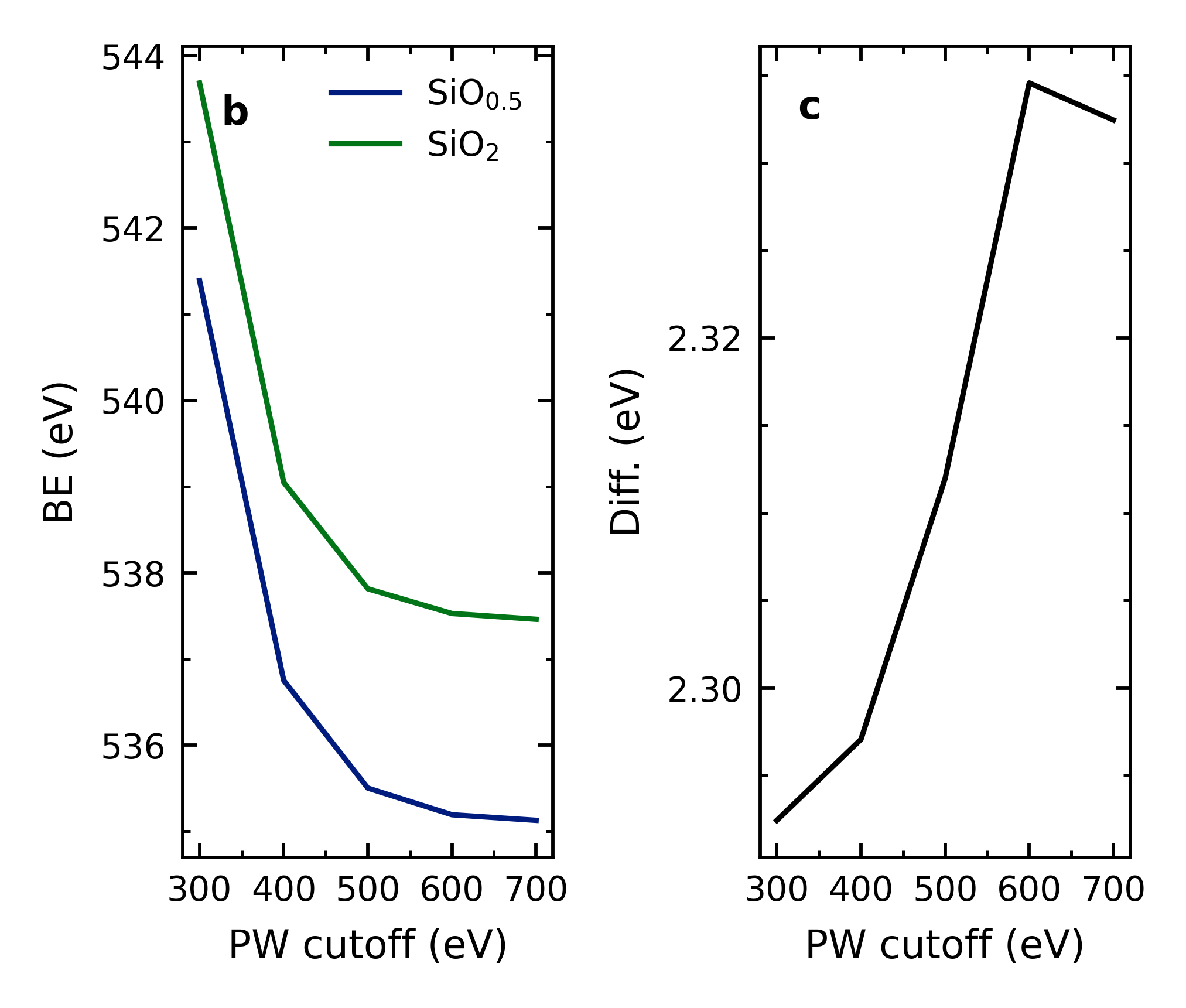}
\caption{\label{fig:escamolecule}C 1s binding energy shifts in the ethyl-trifluoroacetate (\textbf{a}), {\it i.e.} the "ESCA molecule", evaluated as a function of the plane wave (PW) cutoff. Experimental XPS digitized from [O. Travnikova, K. J. Børve, M. Patanen, J. Söderström, C. Miron, L. J. Sæthre, N. M{\aa}rtensson, and S. Svensson, Journal of Electron Spectroscopy and Related Phenomena 185, 191 (2012)] is shown for comparison. Absolute binding energies (\textbf{b}) in SiO$_{0.5}$ \& SiO$_2$ and binding energy shifts (\textbf{c}) between the two SiO$_x$ evaluated as a function of the plane wave cutoff.}
\end{figure}

The binding energy calculation was validated by the C 1s X-ray photoelectron spectrum of ethyl trifluoroacetate, {\it i.e.} the "ESCA molecule". The stick spectrum formed by the calculated core-level binding energies was convoluted with a Gaussian for easier comparison, presented in Figure \ref{fig:escamolecule}(a). In addition, convergence of the results with respect to the plane-wave cutoff are presented for O 1s binding energies of SiO$_2$ and SiO$_{0.5}$ in Figure \ref{fig:escamolecule}(b) using one site in each system. 

\begin{figure}
    \centering
    \includegraphics[width=1\linewidth]{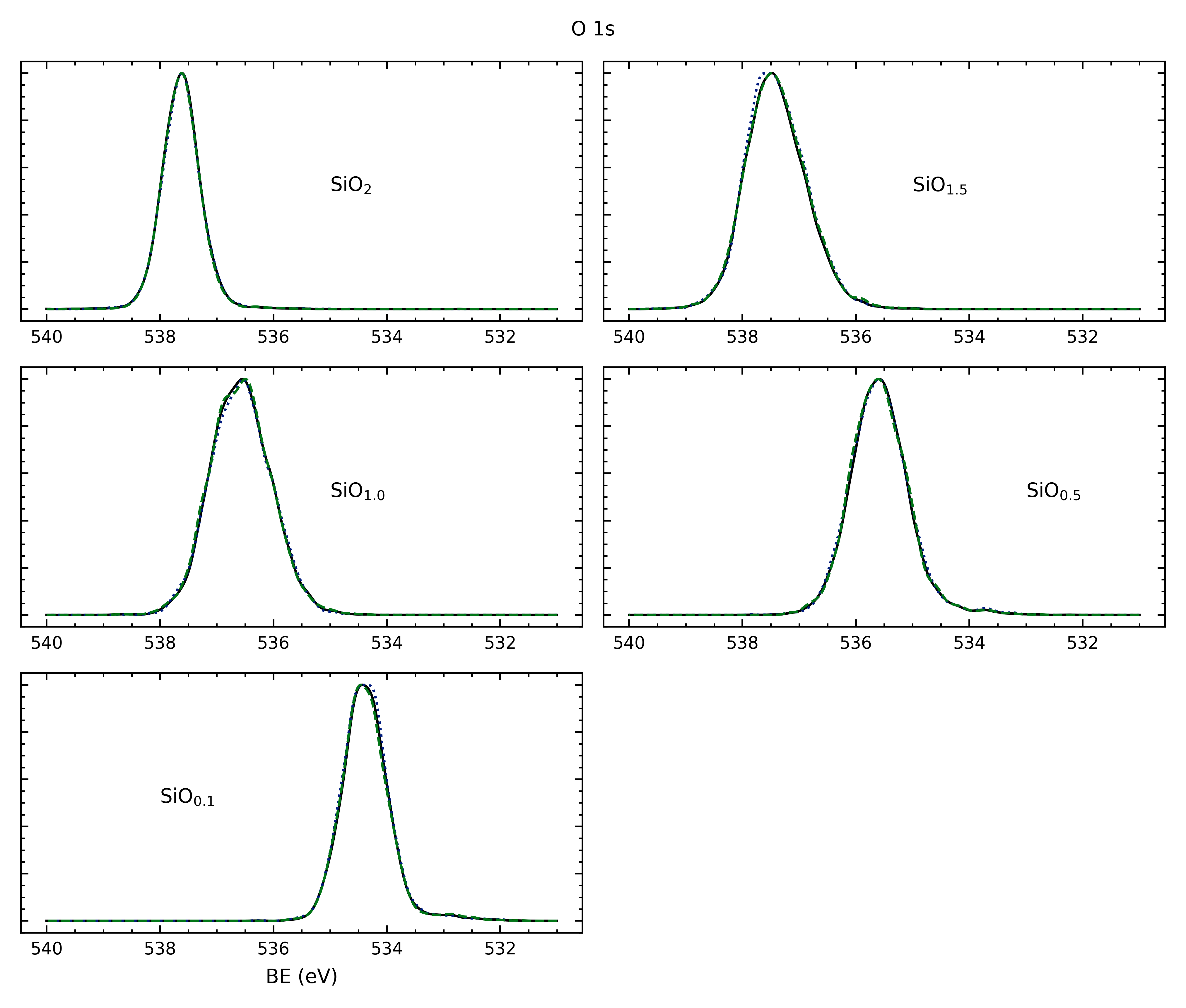}
    \caption{The black solid line shows the resulting O 1s XPS spectrum for each SiO$_x$ system from 5000 calculated binding energies broadened with 0.5 eV FWHM Gaussians . The two dashed lines in color show spectra from 2500 BEs each.}
    \label{fig:o1s-tilastol}
\end{figure}

\begin{figure}
    \centering
    \includegraphics[width=1\linewidth]{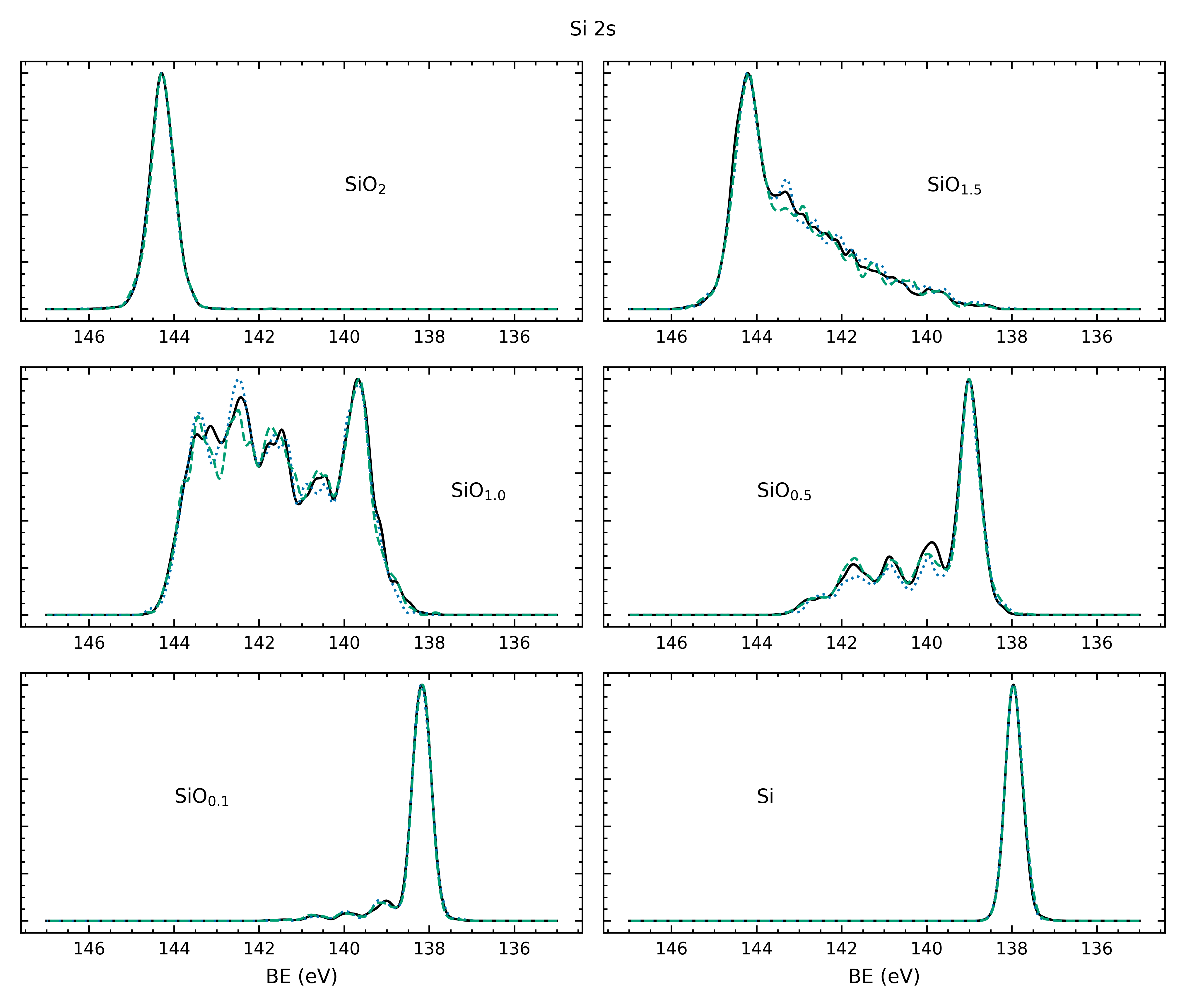}
    \caption{The black solid line shows the resulting Si 2s XPS spectrum for each SiO$_x$ system from 5000 calculated binding energies broadened with 0.5 eV FWHM Gaussians. The two dashed lines in color show spectra from 2500 BEs each.}
    \label{fig:si2s-tilastol}
\end{figure}

The binding energy calculations were further tested using atomistic sites from pure silicon, silicon dioxide and the binding energy shift between them. Firstly, to test the sensitivity of the calculations to the exchange-correlation functional, the same sites were calculated with LDA and PBE, with only 0.01 eV difference in the binding energy shift. DFT structure optimization had a fairly limited effect on the binding energies as well, again only 0.01 eV. 2$\times$2$\times$2 $\Gamma$-centered Monkhorst-Pack k-point sampling produced the same values to $\Gamma$-point calculations within two decimal places for both absolute values and the shift. The effect of cell size was tested with a 2$\times$2$\times$2 supercells, which increased the shift by about 4\%. The results of these checks have been collected in Table \ref{tab:tarkastukset}. To check the convergence of our calculation of 3000 binding energies per system, the spectra have been plotted in Figs. \ref{fig:o1s-tilastol} for the O 1s and \ref{fig:si2s-tilastol} for Si 2s, with two comparison spectra with randomly chosen 1500 BEs for each system.

\par

\begin{table}[h]
    \centering
    \setlength{\tabcolsep}{10pt}
    \begin{tabular}{c c c c c c}
        \hline
        Cell size (\AA) & XC & Other & Si BE (eV) & SiO$_2$ BE (eV) & Shift: BE$_{\text{SiO}_2}$ - BE$_{\text{Si}}$ (eV) \\
        \hline
        15  & LDA & - & 137.45 & 143.28 & 5.83 \\
        15 & LDA & 2x2x2 kpts & 137.45 & 143.28 & 5.83 \\
        15 & PBE & - & 138.17  & 143.99  & 5.82 \\
        15 & PBE & Struct. opt. & 138.23 & 144.06 & 5.83 \\
        30 & PBE & - & 138.18 & 144.26 & 6.08 \\
        \hline
    \end{tabular}
    \caption{Si 2s binding energy for a single site in pure Si \& SiO$_2$ compared with regards to cell size, k-point sampling, xc-functional and whether structure optimization has been performed.}
    \label{tab:tarkastukset}
\end{table}


\FloatBarrier
\section{Population analysis}

\begin{figure}
    \centering
    \includegraphics[width=0.8\linewidth]{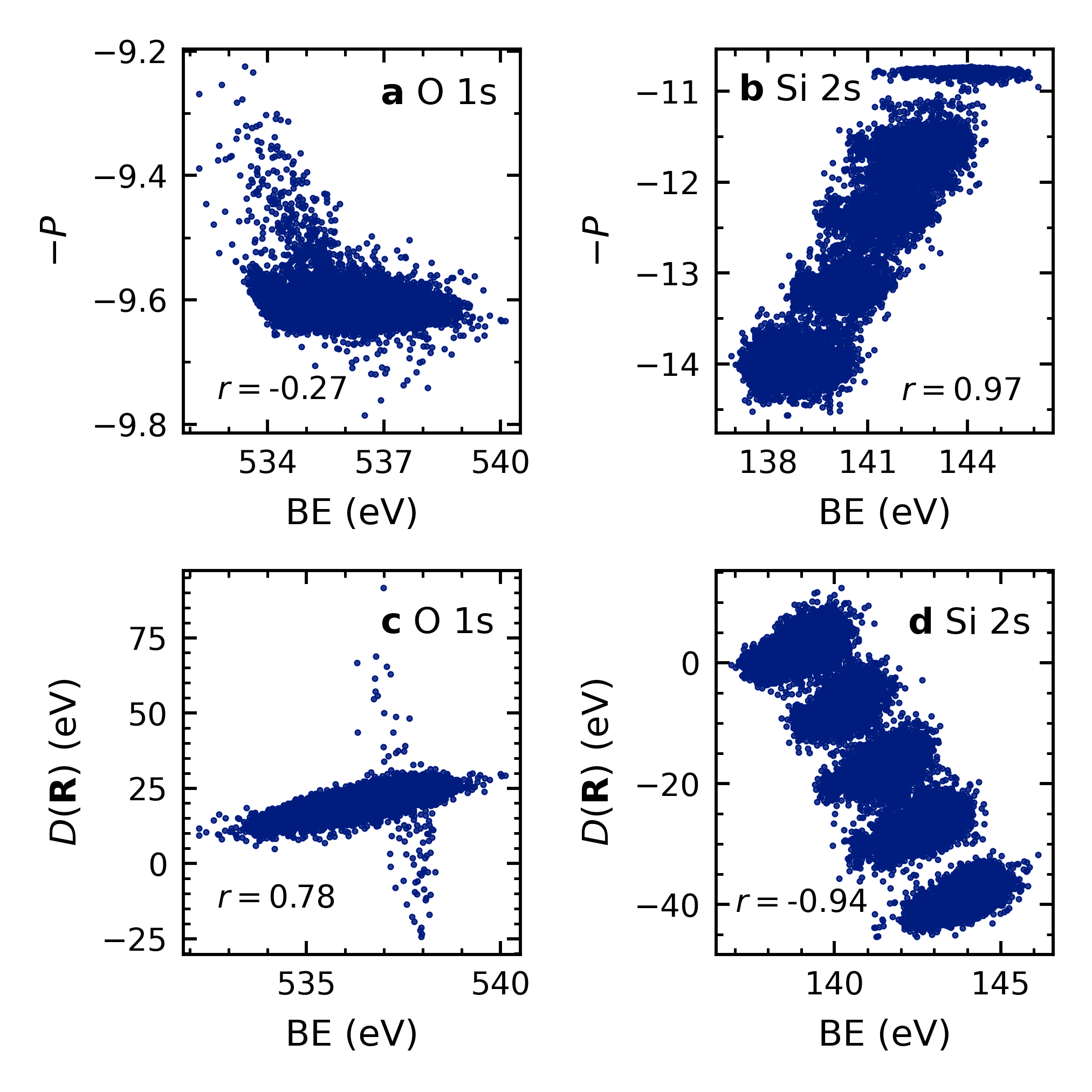}
    \caption{The correlation between the calculated Hirshfeld charge population $P$ and the calculated binding energy for (a) O 1s and (b) Si 2s. A local charge environment based descriptor $D(\mathbf{R})=\sum_i \frac{Z_i - P_i}{r_i}$ is plotted against the calculated binding energy for (c) O 1s and (d) Si 2s. The descriptor consists of distance-weighted ($r_i$) summed charges that are based on the Hirshfeld populations $P_i$ and the atomic number $Z_i$ of neighbor atoms $i$.}
    \label{fig:hirshfeld}
\end{figure}

We have calculated the charge analysis in the ground state for our SiO$_x$ cells. Figures \ref{fig:hirshfeld}a and b depict the respective evaluated Hirshfeld electronic populations against the binding energy at the site. Figures \ref{fig:hirshfeld}c and d show the binding energy and a simple Coulombic descriptor modelling the charge at surrounding each site using the according atomic number. Because of the point-charge nature of the analysis, combining the two rows into a commensurate number is complicated.  

The figure shows that the local charge variation within oxygen atoms is small, and that the charge in its environment gives rise to the core-level binding energy shift. On the contrary, for silicon charge transfer can be more drastic. Due to the proximity of a local valence, this leads to the binding energy shift in silicon atoms, although a competition of positive site-wise and negative local-environment mechanisms is evident. The core-level shift is a outcome of these effects in both the ground state and the core-hole state, most adequately captured by the $\Delta$-SCF calculation employed in the Letter.

\FloatBarrier
\section{Molecular dynamics melt-quench}

\begin{figure}
    \centering
    \includegraphics[width=0.49\linewidth]{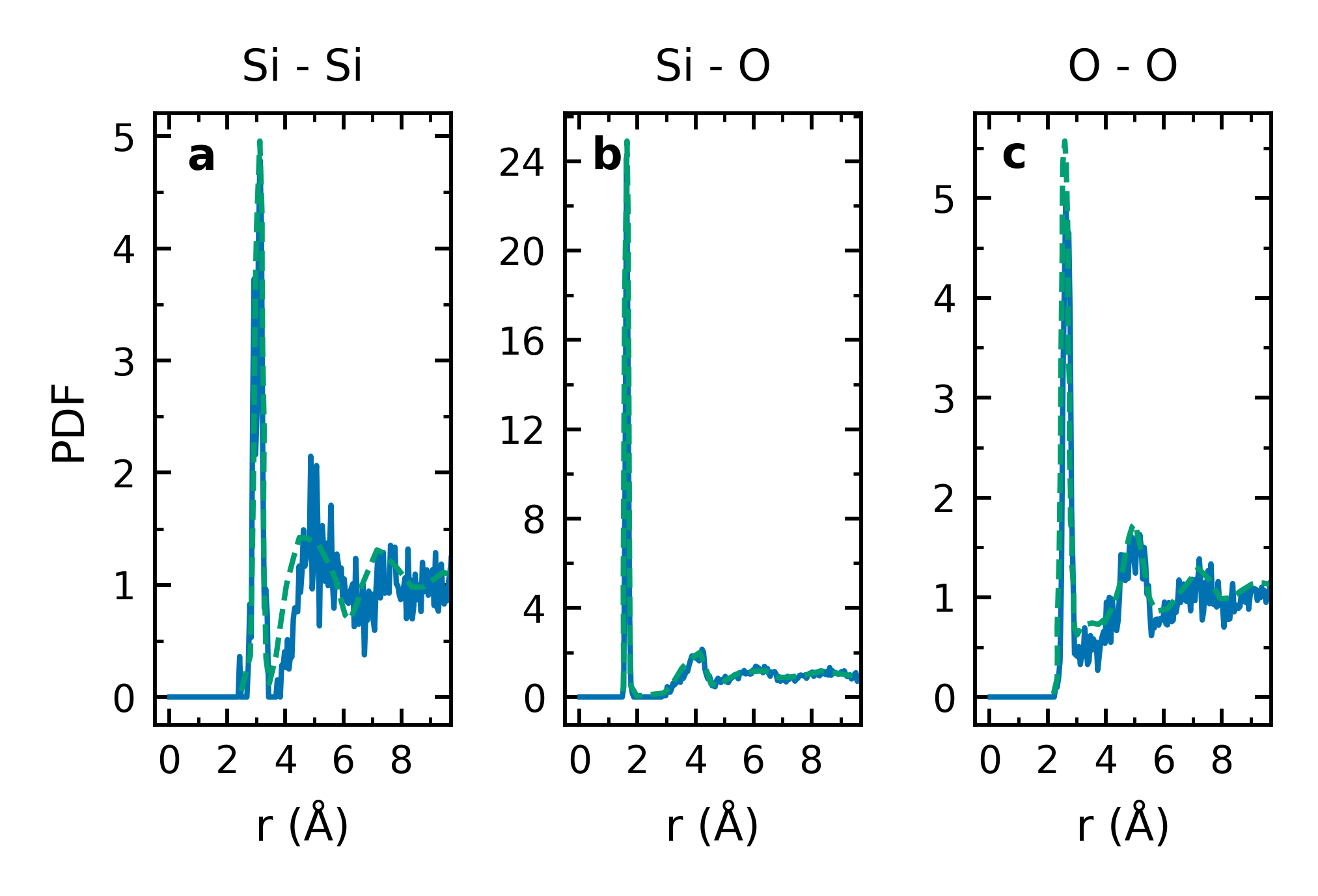}
    \includegraphics[width=0.49\linewidth]{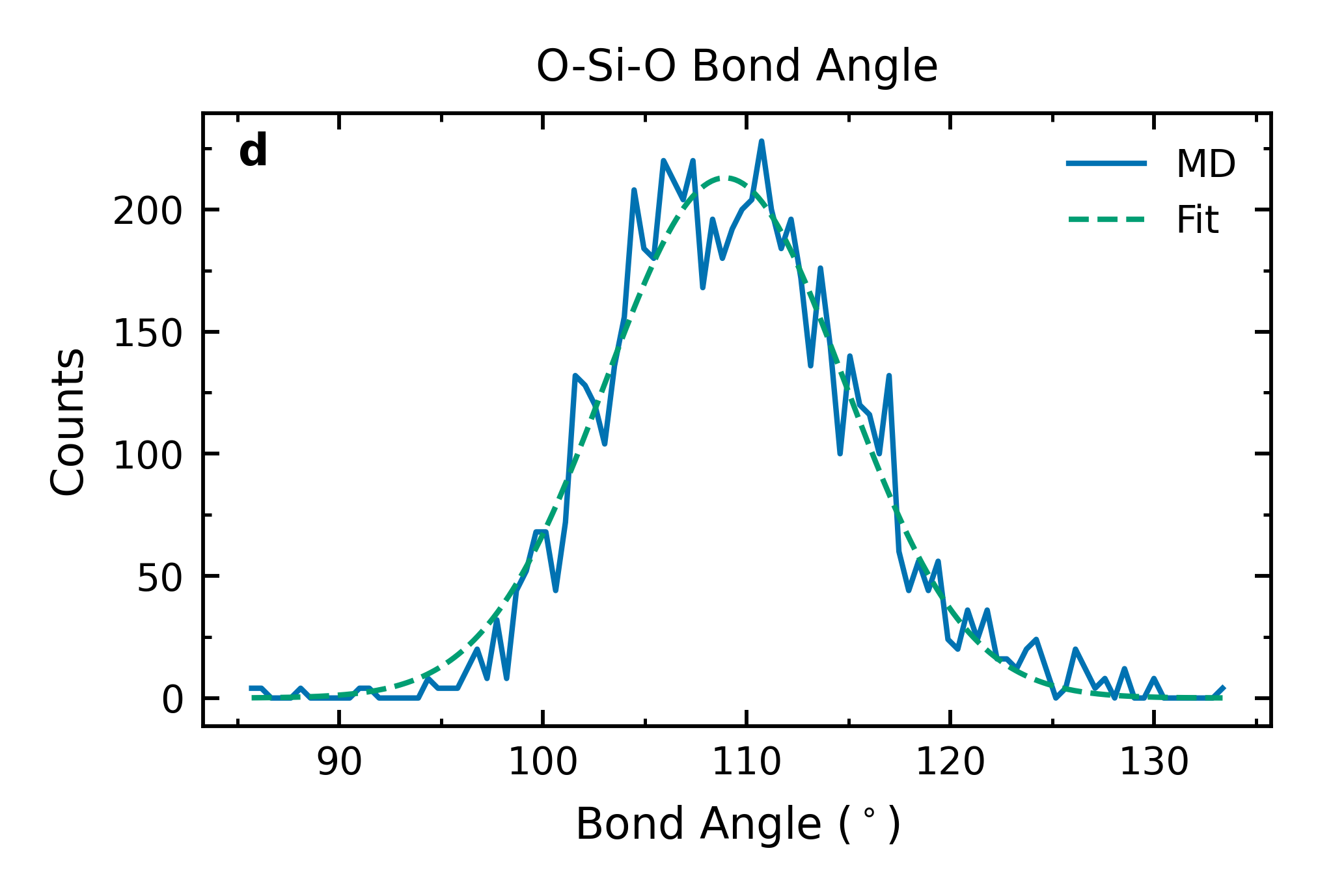}
    \caption{Partial distribution functions for Si-Si, Si-O, and O-O atom pairs (\textbf{a,b,c}) and O-Si-O bond angle distribution (\textbf{d}) for one snapshot of our SiO$_2$-system at 300 K. The dashed line show a comparison to literature PDF [M.-H. Du, A. Kolchin, and H.-P. Cheng, The Journal of Chemical Physics 120, 1044–1054 (2004)] in \textbf{a, b, c}. A fitted gaussian is shown in with dashed line in \textbf{d} with peak center at around 109$^\circ$. The Si-O-Si also gives a reasonable distribution with peak center at around 141$^\circ$}
    \label{fig:siox-pdf}
\end{figure}

While the cooling rate in simulations is orders of magnitude higher than in experiments, reasonable simulated amorphous silicon oxide has been generated in literature with cooling rates in the range of $10^{11} - 10^{15}$ K/s [D. J. Cole, M. C. Payne, G. Cs{\'a}nyi, S. Mark Spearing, and L. Colombi Ciacchi, The Journal of Chemical Physics 127, 10.1063/1.2799196 (2007); J. Sarnthein, A. Pasquarello, and R. Car, Physical Review B 52, 12690–12695 (1995); M. D. Kluge, J. R. Ray, and A. Rahman, Physical Review B 36, 4234–4237 (1987)]. To test the short-range order, partial radial distribution functions and bond angle distributions of our simulations have been compared to literature [M.-H. Du, A. Kolchin, and H.-P. Cheng, The Journal of Chemical Physics 120, 1044–1054 (2004);  S. S. Jena, S. Singh, and S. Chandra, Applied Physics A 129, 10.1007/s00339-023-07020-2 (2023)] with a reasonable match (Fig. \ref{fig:siox-pdf}). 

\begin{figure}
    \centering
    \includegraphics[width=0.49\textwidth]{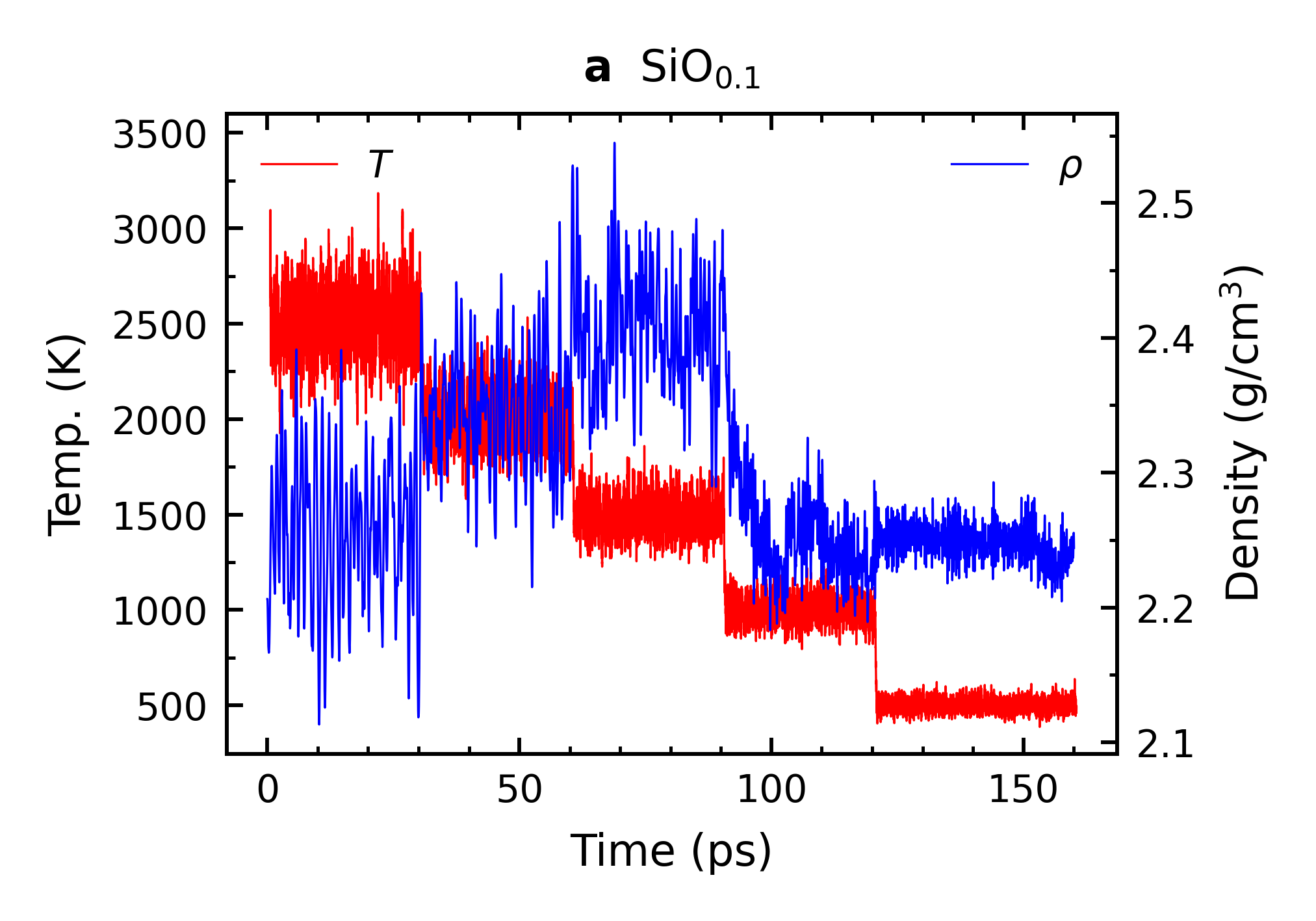}
    \includegraphics[width=0.49\textwidth]{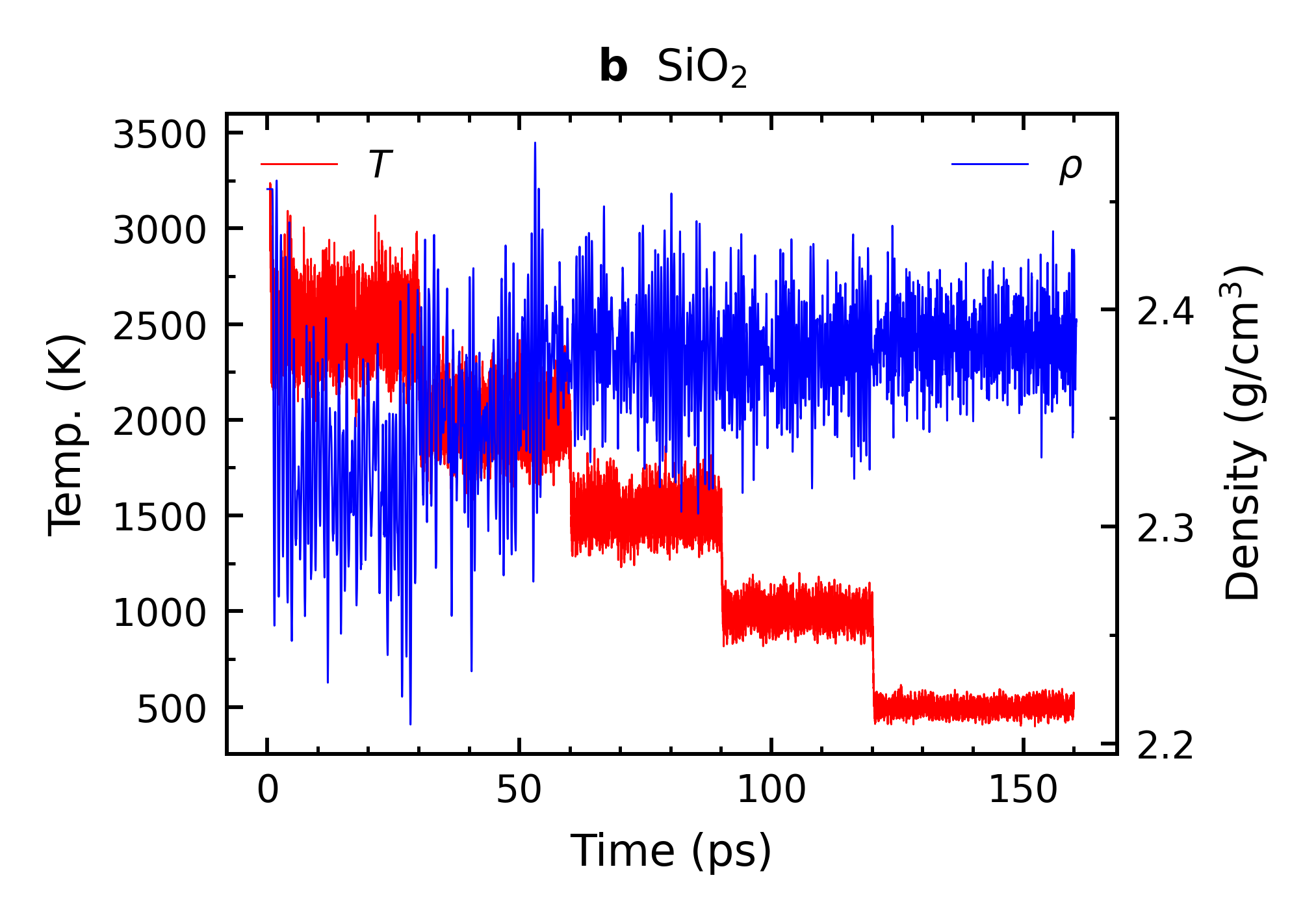}
    \caption{Temperature and density during the NpT cooling phase of the melt-quench of the two extremes of our SiO$_x$-compositions.}
    \label{fig:time-temp+rho}
\end{figure}

Densities after cooling for our SiO$_2$ cells end up at around 2.18 - 2.20 g/cm$^3$ at 300 K, which matches experimental density of 2.2 g/cm$^3$. For pure Si, the density ended up at $2.25 - 2.26$ g/cm$^3$, which is also reasonable compared to experimental density (2.3 g/cm$^3$). 
Fig. \ref{fig:time-temp+rho} shows how temperature and density behave as a function of time for one cooling run for SiO$_{0.1}$ and SiO$_2$. Temperature fluctuates around the target temperature of the cooling schedule. We can see lower densities with higher temperatures as expected and it approaches the expected room‑temperature density. 

\begin{figure}
\begin{minipage}{0.49\linewidth}
        \includegraphics[width=\linewidth]{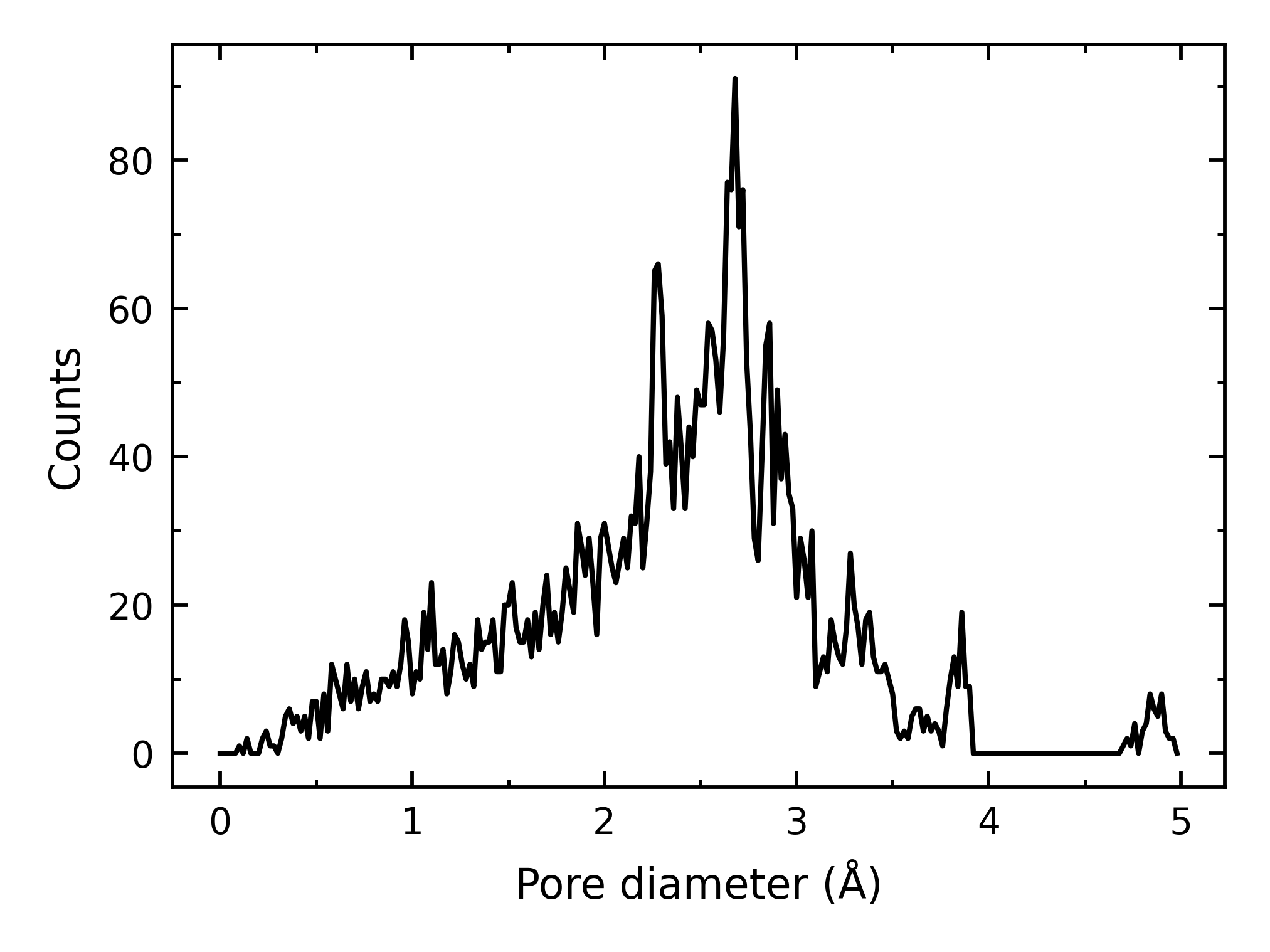}
        \caption{Void size distribution curve for SiO$_2$.}
        \label{fig:voidsizedistr}
\end{minipage}\hfill
\begin{minipage}{0.49\linewidth}
        \includegraphics[width=\textwidth]{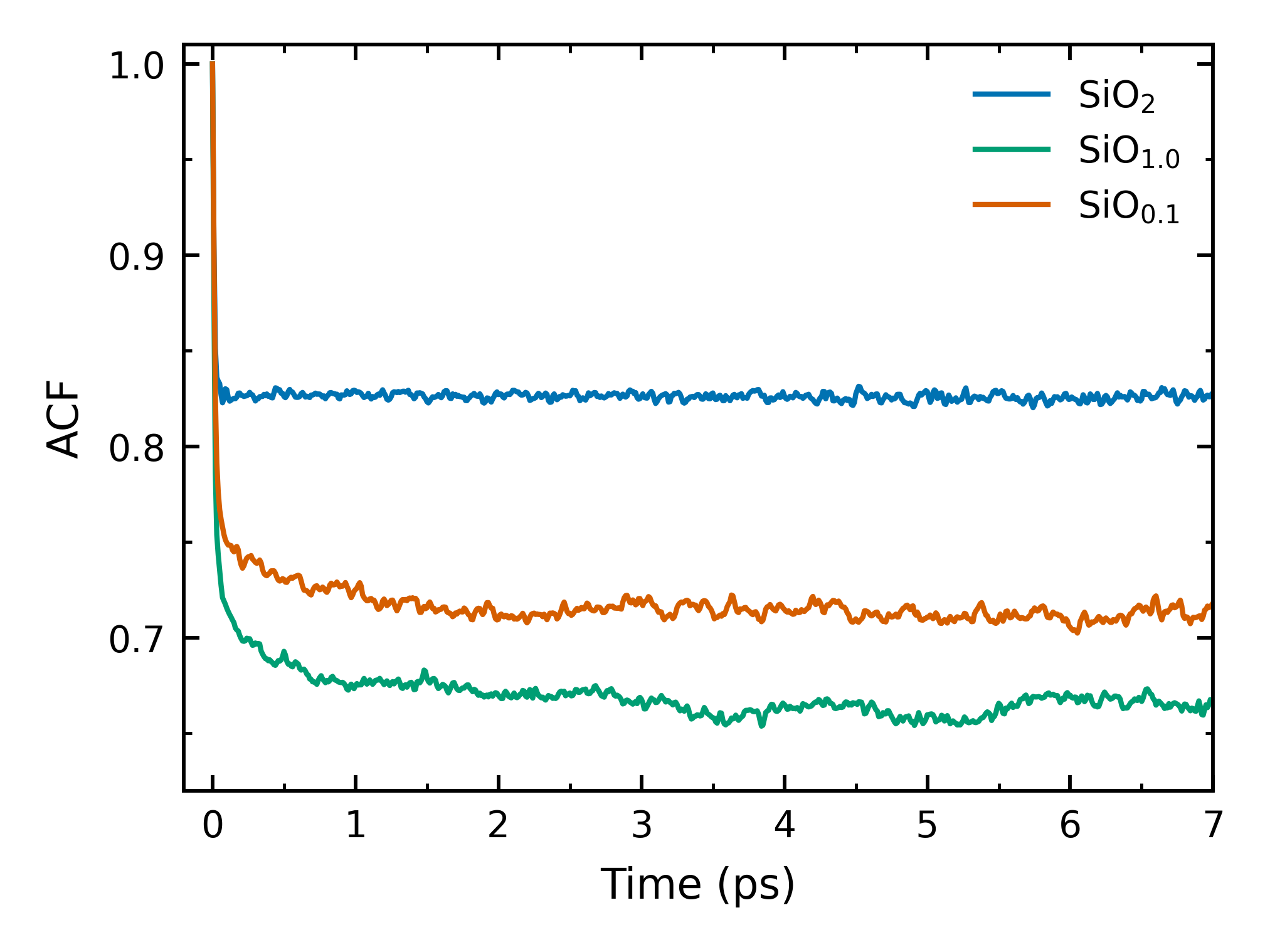}
        \caption{Example autocorrelation functions for SiO$_x$.}
        \label{fig:melt-acf}
\end{minipage}
\end{figure}

Structurally, our SiO$_2$ cells mostly consist of the typical tetrahedra of 1 Si and 4 O atoms with some defects and unevenness in the distribution of atoms. Uneven distribution in SiO$_2$ cells was quantified with MOF explorer [M. Usdin, G. Chung, and R. Snurr, MOF Explorer, https://mausdin.github.io/MOFsite/mofPage.html (2015)] by calculating and comparing the void distribution curve (Fig. \ref{fig:voidsizedistr}), which gives us some insight into the medium-range order/spatial heterogeneity. The peak in the curve falls between void diameters of around $2-3$ Å, which is comparable to curves for DFT melt-quenched SiO$_2$ by Jena et al. [S. S. Jena, S. Singh, and S. Chandra, Applied Physics A 129, 10.1007/s00339-023-07020-2 (2023)]. 
Since the amorphous structure features large amount of randomness, particularly for substoichiometric compositions, we have calculated several annealing branches for each composition. The autocorrelation of the LMBTR-descriptor of the local atomistic environment (Fig. \ref{fig:melt-acf}) showed that after a few picoseconds the simulations are as statistically independent as they will be. Based on this, we branch out into a cooling run every three ps from the main melt run endpoint.

\FloatBarrier
\section{Experimental}
\begin{figure}
    \centering
    \includegraphics[width=1\linewidth]{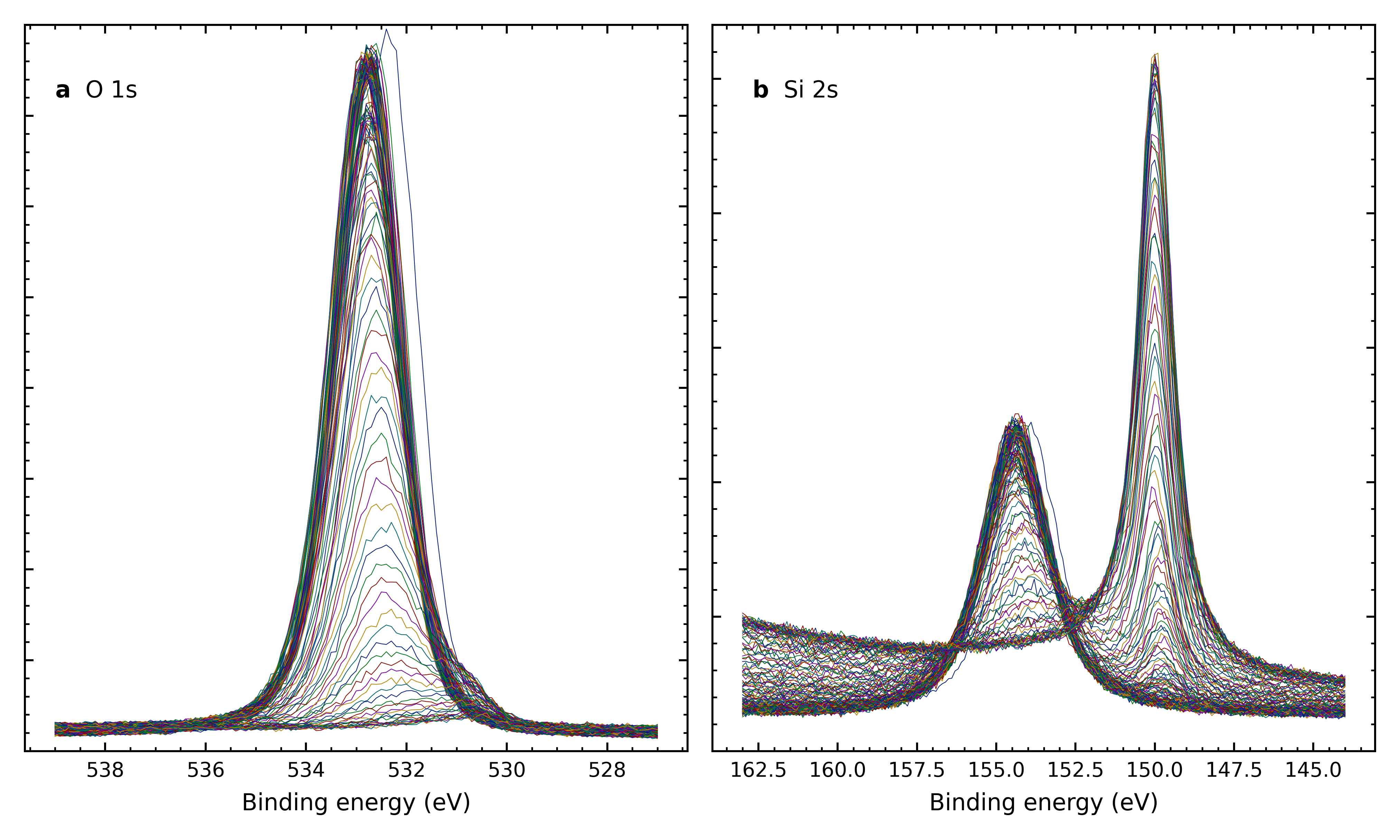}
    \caption{Spectra from a 40 nm silicon oxide sample sputtered with a monoatomic 300 eV argon ions 20 s/cycle. Representative spectra have been selected for Figure 2 in the Letter.}
    \label{fig:kaikkispektrit}
\end{figure}

The raw sputtered XPS data has been presented in Figure \ref{fig:kaikkispektrit}. The layer-wise spectra have been estimated from these following the method described in the Letter and representative ones have been chosen to be plotted in Fig. 2 of the Letter.

\begin{figure}
    \centering
    \includegraphics[width=0.5\linewidth]{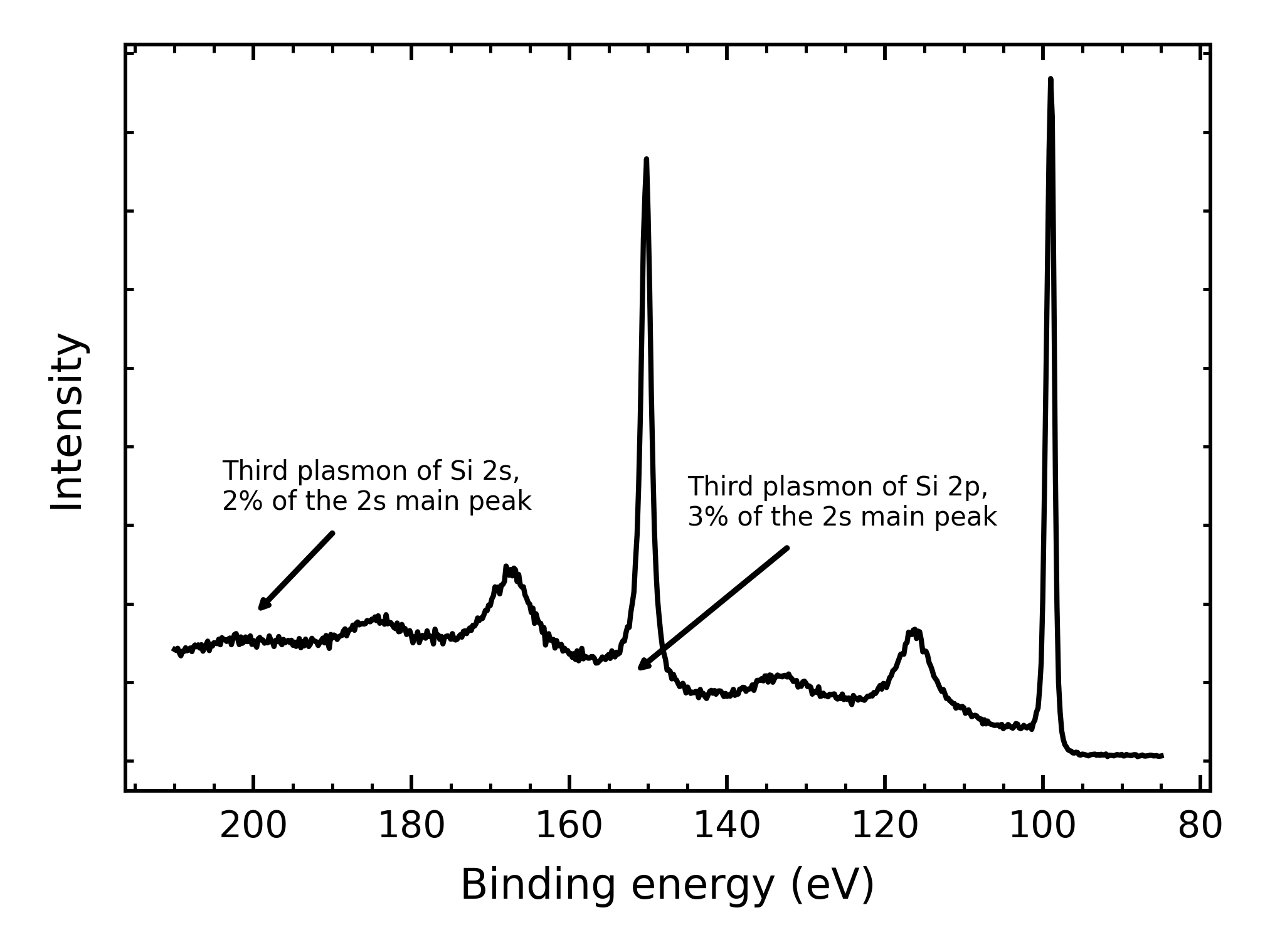}
    \caption{The third plasmon of Si 2p is underneath the Si 2s peak. The intensity of the third plasmon peak of Si 2s is 2\% of the intensity of the 2s main peak. Based on this and the intensity ratio between Si 2p and Si 2s, we estimate the intensity of the third Si 2p plasmon under the Si 2s peak to be 3\% relative to 2s peak.}
    \label{fig:plasmon}
\end{figure}

The third plasmon peak of Si 2p overlaps with our Si 2s peak. At the most intense case (shown in Figure \ref{fig:plasmon}), we find the third plasmon loss peak of Si 2s to be about 2\% of the Si 2s main peak. Assuming comparable relative plasmon intensities, we scale this by the intensity ratio $\frac{I_{\text{Si 2p}}}{I_{\text{Si 2s}}}$, which gives us an estimate that the third Si 2p plasmon peak under the Si 2s main peak is around 3\% relative to the Si 2s main peak intensity. We treat this intensity as negligible in the context of this work.

\begin{figure}
    \centering
    \includegraphics[width=0.5\linewidth]{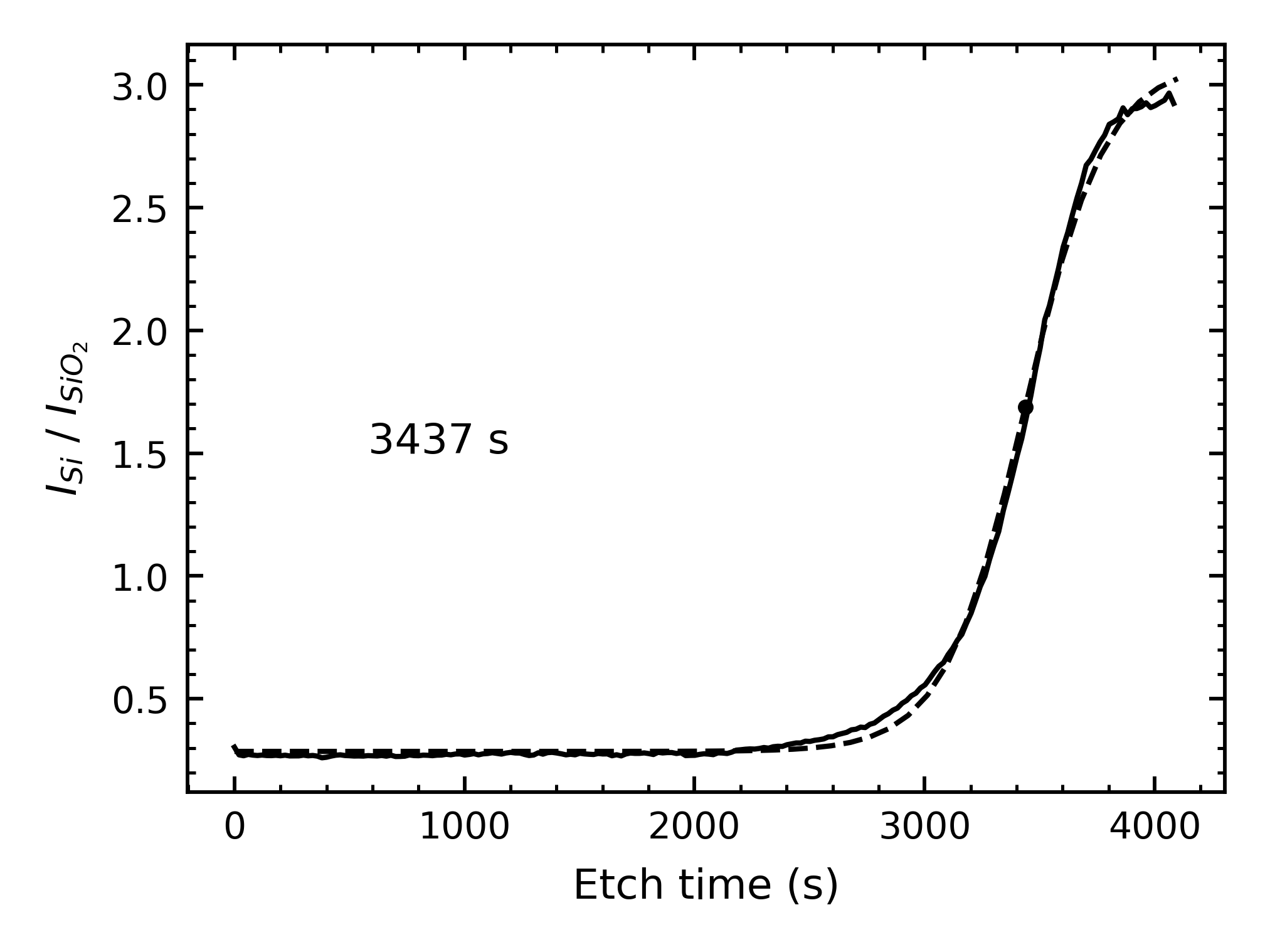}
    \caption{Ratio of the integrated intensity between the SiO$_2$ and Si peaks as a function of sputtering time. A fitted sigmoid
function midpoint gives an estimation for time it takes to sputter the oxide layer.}
    \label{fig:thickness}
\end{figure}

To estimate the sputter rate, we investigate the intensity ratio between the SiO$_2$ and Si components in Si 2s. The ratio as a function of sputter time is shown in Fig. \ref{fig:thickness}. From a fitted sigmoid function center point, we get a sputter time of 3434 s to sputter the oxide layer. Based on this and the ellipsometrically determined sample thickness of 40 nm, the sputter rate is 0.01 nm/s.

\section{Inelastic mean free path (IMFP)}

\begin{figure}
    \centering
    \includegraphics[width=0.5\linewidth]{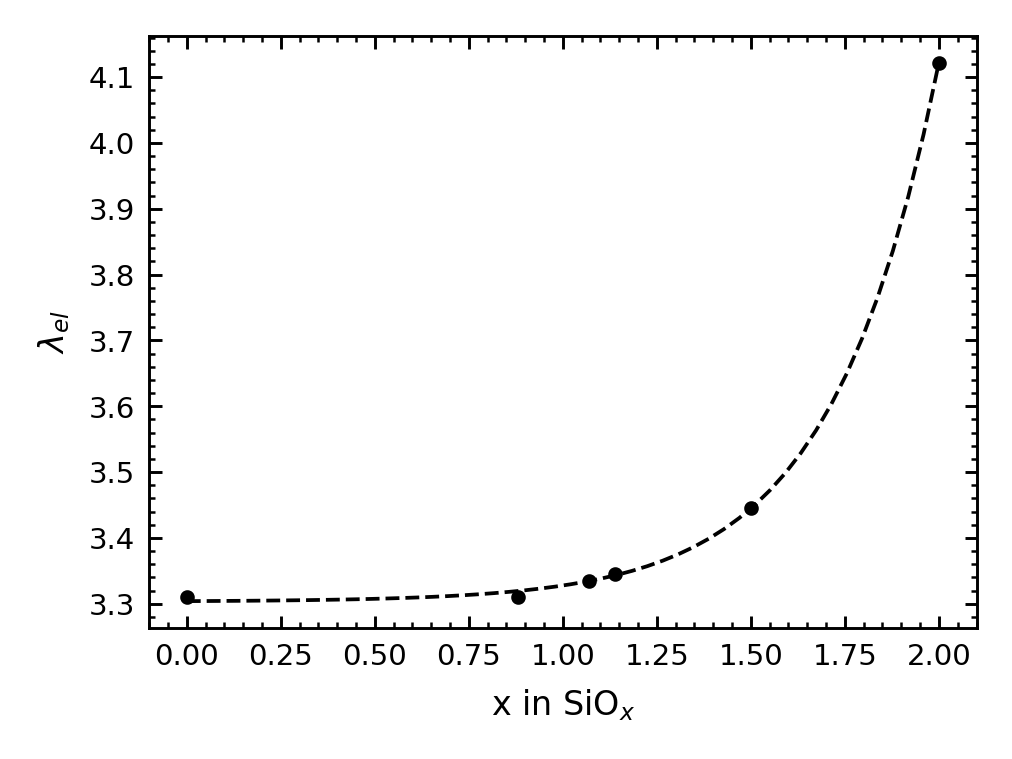}
    \caption{Mean free path $\lambda_{el}$ as a function of x in SiO$_x$}
    \label{fig:x-lambdael}
\end{figure}

The following investigation was done with a previous measurement set of depth resolved XPS, but the conclusion of this sections should hold for the current measurement and its analysis as well. Fig. \ref{fig:x-lambdael} shows how IFMP varies with x in SiO$_x$. Values are calculated with the TPP-2M formula \cite{Tanuma1994}, with band gap values for substoichiometric SiOx from [V. Kruchinin, T. Perevalov, G. Kamaev, S. Rykhlitskii, V. Gritsenko, Optics and Spectroscopy 127, 5 (2019)].

\begin{figure}
    \centering
    \includegraphics[width=0.5\linewidth]{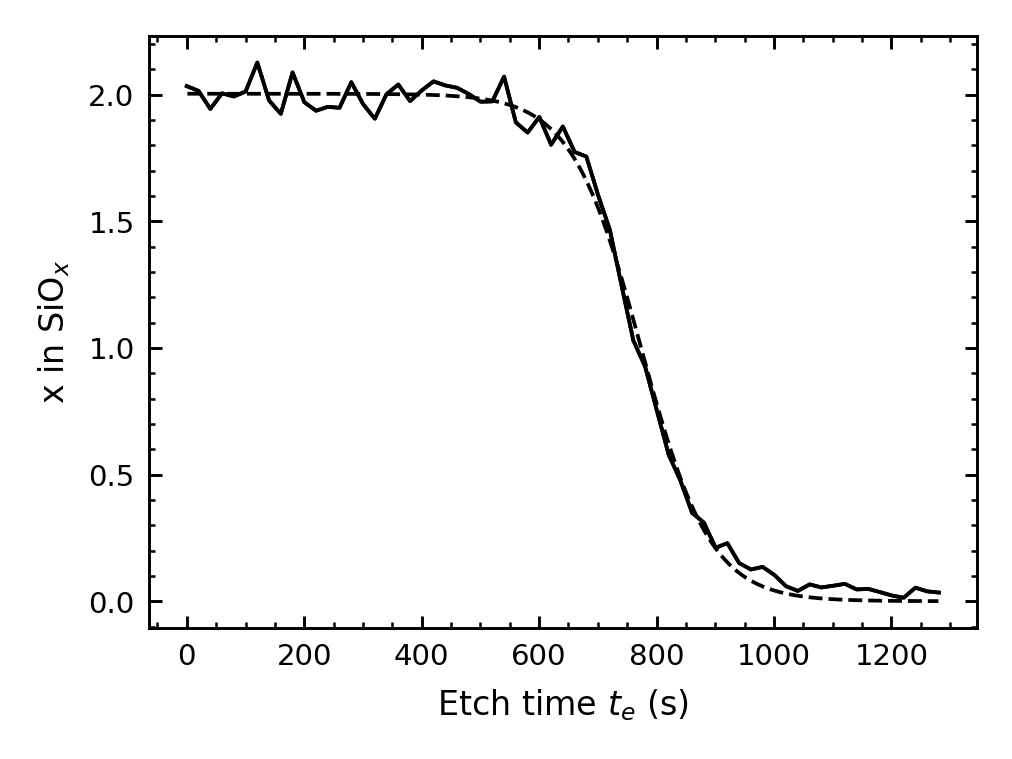}
    \caption{x in SiO$_x$ as a function etch time $t_e$}
    \label{fig:etchtime-x}
\end{figure}

In Fig. \ref{fig:etchtime-x}, x in SiO$_x$ as a function of etch time is estimated from $x = \frac{I_{O1s}}{I_{Si2s}} \cdot K$, here $I_{O1s}, I_{Si2s}$ are the integrated areas under the O1s and Si2s peaks after background removal. Sensitivity ratio $K$ estimated from the first few spectra with the assumption that x=2 with them.

\begin{figure}
    \centering
    \includegraphics[width=0.5\linewidth]{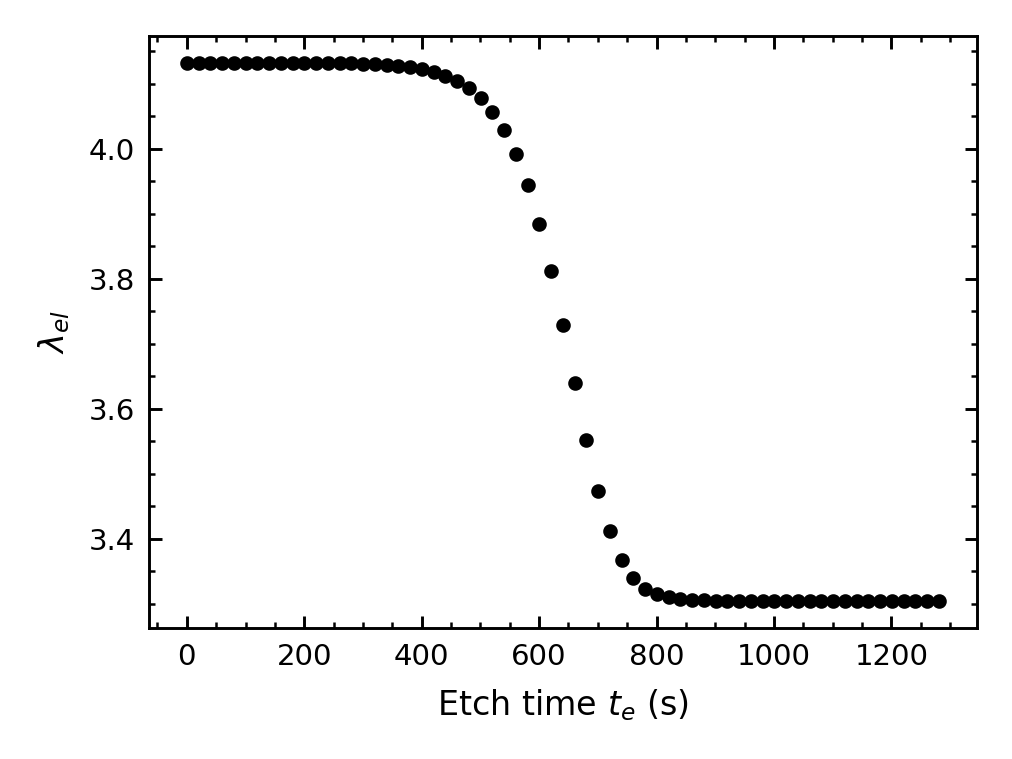}
    \caption{Mean free path $\lambda_{el}$ as a function etch time $t_e$}
    \label{fig:etchtime-lambdael}
\end{figure}

Based on the previous results, the free mean path as a function etch time is calculated for Fig. \ref{fig:etchtime-lambdael}. 

The reconstructed spectra (Fig. 2) calculated with varying $\lambda_{el}$ features very limited difference to the one calculated with constant $\lambda_{el}$. We have therefore decided to use the simpler approach, a constant mean free path, with a value from the average of the two extremes, Si and SiO$_2$.

\FloatBarrier
\section{Calculated spectra}

\begin{figure}
    \centering
    \includegraphics[width=\linewidth]{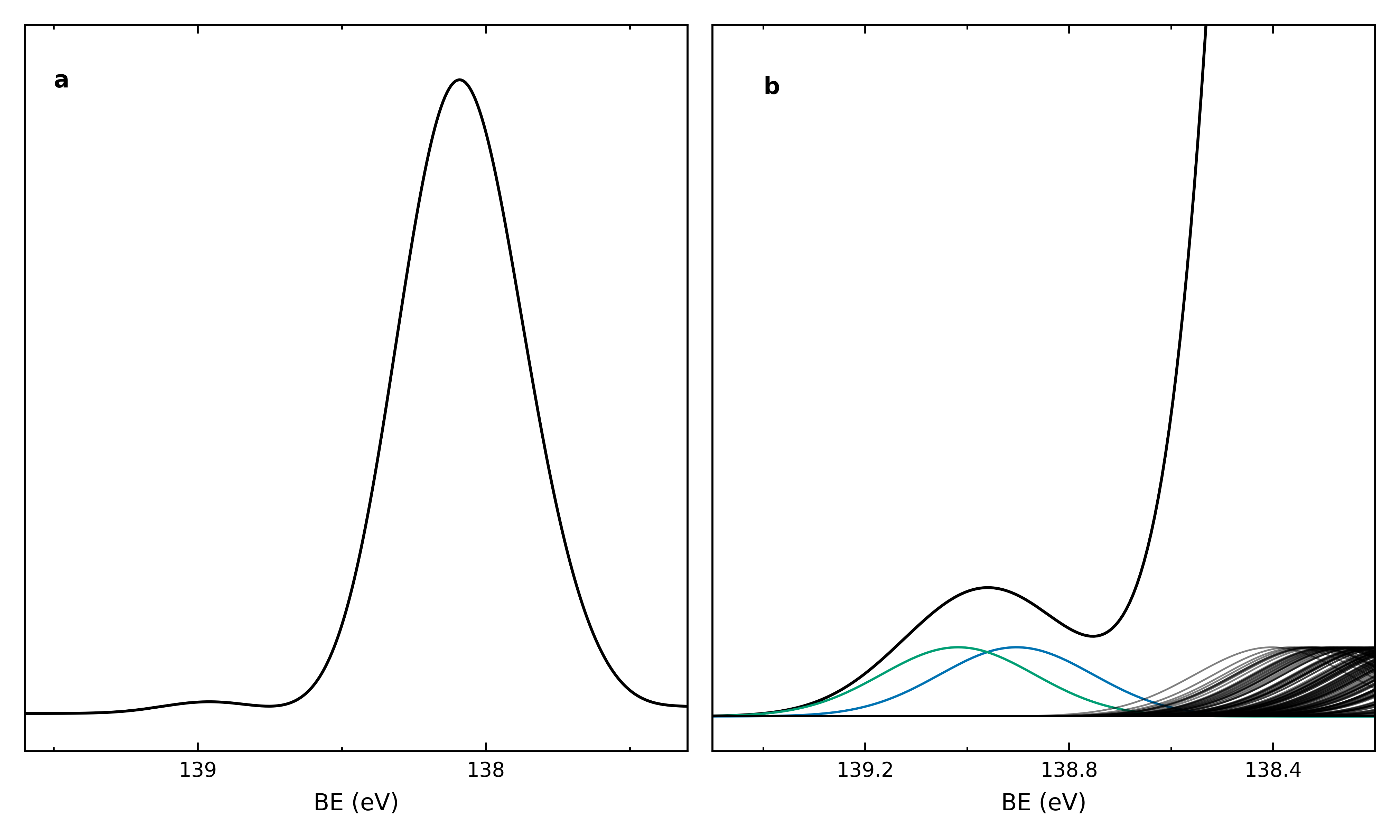}
    \caption{\textbf{a}: The calculated Si 2s spectrum for a simulation cell with one oxygen atom and 149 silicon atoms. \textbf{b}: Zoomed in plot of the spectrum shows the two silicon atoms, whose binding energies have clearly shifted from the main peak.}
    \label{fig:sidepeak}
\end{figure}

To investigate spectra at low oxygen concentrations, a simulation cell with 1 oxygen and 149 silicon atoms (SiO$_{0.0067}$) was calculated with the same procedure: melt-quench MD and $\Delta$-SCF for each Si atom site. As shown in Fig 2 in the Letter, with Si 2s in SiO$_x$ the peaks shift as $x$ increases. The main peak for the simulation cell with one oxygen consists of binding energies typical for pure Si but it features two atoms, whose binding energies have shifted towards higher binding energies. The calculated spectrum is shown in Figure \ref{fig:sidepeak}(a) and (b) shows a zoomed plot of the side peak and the two Gaussian-broadened binding energies that contribute to it.

\FloatBarrier
\section{Reconstruction of layer-wise spectra}
\begin{figure}
    \centering
    \includegraphics[width=0.99\linewidth]{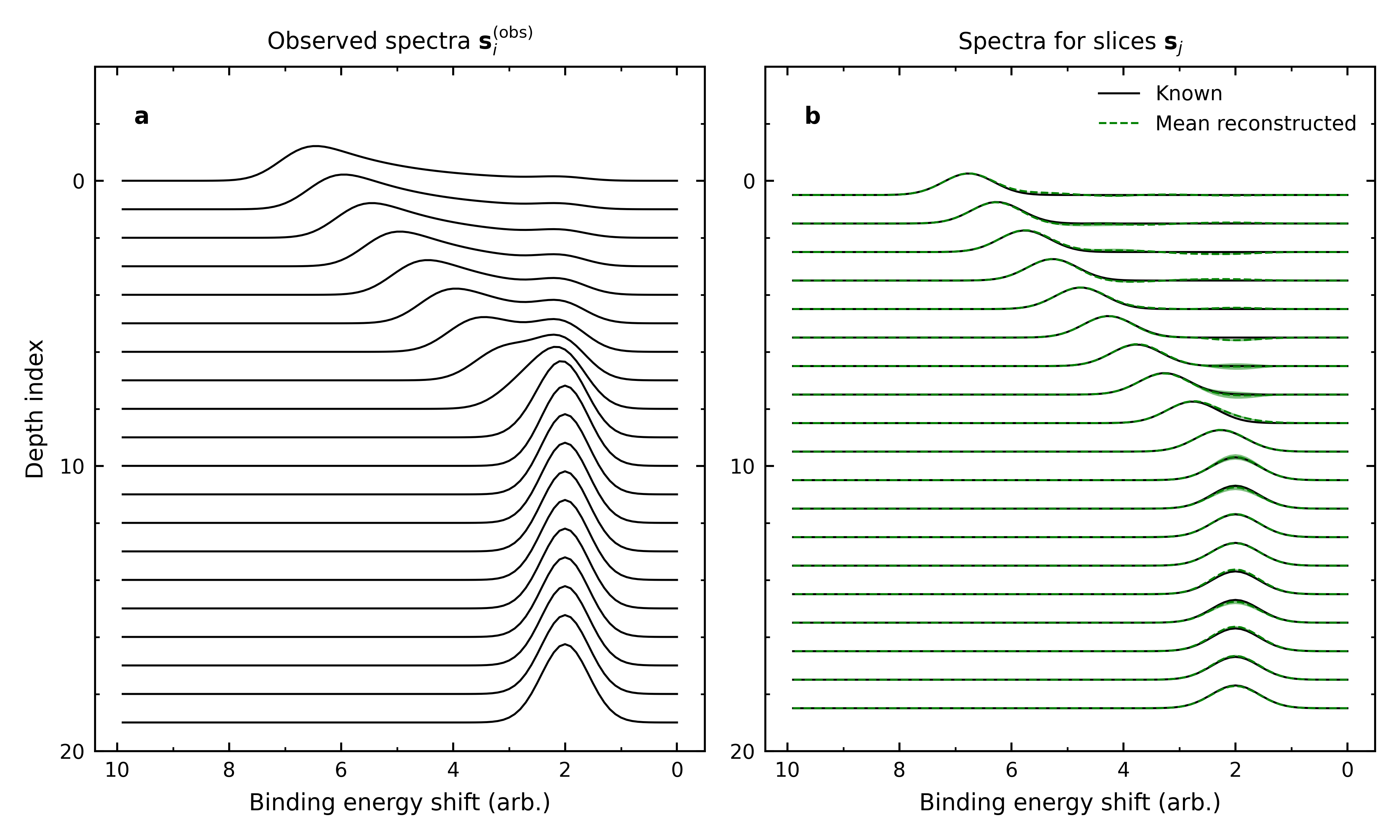}
    \caption{\label{fig:rec}Validation of the reconstruction method using synthetically generated test data with binding-energy axis in arbitrary units and depth shown as the slice index. The depth-integrated sputtered spectra with added noise are shown in \textbf{a}. The known spectra for layers and their reconstruction from the integrated sputtered data are shown in \textbf{b}. The sputter $\Delta r/\lambda_\mathrm{el}$ ratio of the data was 1/3. The shading shows the min-max range in a 1000-fold bootstrap procedure with values $\pm$20~\%. The reconstructed layers are drawn vertically offset to highlight their correspondence with the layer removed in a sputtering cycle.}
\end{figure}

We validated the reconstruction method for the layer-wise spectra using a known test case. We defined a spectrum $\mathbf{s}_r$ at a given depth $r\geq0$ as a Gaussian function $\mathbf{s}_r(E) = \mathrm{e}^{-\frac{1}{2}(E-E_r)^2/\sigma^2},$
where $E$ is the binding energy and $\sigma\approx\mathrm{FWHM}/2.35$ of the peak, evaluated on a grid. The peak position $E_r$ shifts linearly to depth $r=10$, remaining constant deeper in the sample. Each sputtering exposure removes a layer with the thickness $\Delta r$ from the surface of the system, and therefore the integrated signal after sputtering $i$ cycles reads
\begin{equation}
    \mathbf{s}_i^{\mathrm{(obs)}}(E) = \int_{i\Delta r}^\infty \mathrm{e}^{-(r'-i\Delta r)/\lambda_\mathrm{el}}\mathbf{s}_{r'}(E)\mathrm{d}r'
\end{equation}
reflecting the fact that all sample below this new surface layer contributes to the observed signal. Analogously, the signal $\mathbf{s}_j(E)$ from a layer removed during $j$th sputtering cycle is
\begin{equation}
    \mathbf{s}_j(E) = \int_{(j-1)\Delta r}^{j\Delta r} \mathrm{e}^{-(r'-(j-1)\Delta r)/\lambda_\mathrm{el}}\mathbf{s}_{r'}(E)\mathrm{d}r'.
\end{equation}
We used $\Delta r=1.0$, $\lambda_\mathrm{el}=3.0$ and calculated the integrated spectra numerically using a small $\mathrm{d}r'$. To test the reconstruction method, we applied 1000-fold bootstrap method in which we allowed the inherent $\Delta r/\lambda_\mathrm{el}$ ratio to vary $\pm$20\% from the value 1/3. The spectra integrated over all remaining depth, and over the layers are shown in Figure \ref{fig:rec} together with the reconstructed layer-wise spectra.

\end{document}